\documentclass[12pt]{article}

\usepackage{amsmath}
\usepackage{amssymb}
\usepackage[symbol]{footmisc}
\usepackage{latexsym}
\usepackage{caption}
\usepackage{subcaption}
\usepackage[hyperref]{xcolor}
\usepackage{epsfig}
\graphicspath{{projectplots/}}

\usepackage{tabularx} 
\usepackage{amsmath}  
\usepackage{graphicx} 
\usepackage{cite} 
\usepackage{hyperref} 
\hypersetup{colorlinks,breaklinks,
    colorlinks=true,
    linkcolor=blue,
    filecolor=magenta,      
    urlcolor=teal,
    citecolor=blue}
\def\beq{\begin{equation}}
\def\eeq{\end{equation}}
\def\bea{\begin{eqnarray}}
\def\eea{\end{eqnarray}}

\usepackage{relsize}
\usepackage{indentfirst}
\usepackage{amssymb}
\usepackage{bookmark}
\usepackage{amsthm}
\usepackage{amsmath}

\newcommand{\newc}{\newcommand}
\def\eq$#1${\begin{equation}#1\end{equation}}
\def\gat$#1${\begin{gather}#1\end{gather}}
\def\bal$#1${\begin{align}#1\end{align}}
\def\eqarr$#1${\begin{eqnarray}#1\end{eqnarray}}
\newc{\pa}{\partial}
\newc{\alp}{\alpha}
\newc{\gam}{\gamma}
\newc{\Gam}{\Gamma}
\newc{\del}{\delta}
\newc{\eps}{\epsilon}
\newc{\lam}{\lambda}
\newc{\sig}{\sigma}
\newc{\ups}{\upsilon}
\newc{\ome}{\omega}
\newc{\pphi}{\varphi}
\newc{\nonum}{\nonumber}
\newc{\hami}{\text{\textbf{\lat{H}}}}
\newc{\gren}{\mathcal{G}}
\newc{\lagr}{\mathcal{L}}
\newc{\timor}{\mathcal{T}}
\newc{\prop}{\mathcal{K}}
\newc{\zcal}{\mathcal{Z}}
\newc{\operx}{\text{\textbf{\lat{x}}}}
\newc{\opera}{\text{\textbf{\lat{a}}}}
\newc{\operp}{\text{\textbf{\lat{p}}}}
\newc{\operl}{\text{\textbf{\lat{L}}}}
\newc{\gfv}{g^{(5)}}
\newc{\kfv}{\kappa_{(5)}}
\newc{\tf}{\tilde{f}}
\newc{\tlam}{\tilde{\Lambda}}
\newc{\tl}{\tilde{\lam}}
\newc{\dist}{\displaystyle}
\newc{\ra}{\rightarrow}
\newc{\Ra}{\Rightarrow}

\begin{document}

\begin{titlepage}

\vspace*{1cm}
\begin{center}
{\bf \Large Anti-Gravitating Brane-World Solutions\\[2mm] for a de Sitter Brane
in Scalar-Tensor Gravity}

\bigskip \bigskip \medskip

{\bf Panagiota Kanti},$^{1,}$\footnote[1]{Email: pkanti@cc.uoi.gr}  
{\bf Theodoros Nakas},$^{1,}$\footnote[2]{Email: thnakas@cc.uoi.gr} and
{\bf Nikolaos Pappas}$^{2,}$\footnote[3]{Email: npappas@cc.uoi.gr}

\bigskip
$^1${\it Division of Theoretical Physics, Department of Physics,\\
University of Ioannina, Ioannina GR-45110, Greece}

\bigskip 
$^2${\it Nuclear and Particle Physics Section, Physics Department,\\
National and Kapodistrian University of Athens, Athens GR-15771, Greece}

\bigskip \medskip
{\bf Abstract}
\end{center}
In the context of a five-dimensional theory with a scalar field nonminimally coupled
to gravity, we look for solutions that describe novel black-string or maximally symmetric
solutions in the bulk. The brane line element is found to describe a Schwarzschild--(anti)--de
Sitter spacetime, and, here, we choose to study solutions with a positive four-dimensional
cosmological constant. We consider two different forms of the coupling function of the
scalar field to the bulk scalar curvature, a linear one and a quadratic one. In the linear
case, we find solutions where the theory, close to our brane, mimics an ordinary
gravitational theory with a minimally coupled scalar field giving rise to an
exponentially decreasing warp factor in the absence of a negative bulk cosmological constant.
The solution is characterized by the presence of a normal gravity regime around our brane
and an antigravitating regime away from it. In the quadratic case, there is no 
normal-gravity regime at all; however, scalar field and energy-momentum tensor
components are well defined and an exponentially decreasing warp factor emerges
again. We demonstrate that, in the context of this theory, the emergence of a
positive cosmological constant on our brane is always accompanied by an
antigravitating regime in the five-dimensional bulk.

\end{titlepage}

\renewcommand*{\thefootnote}{\arabic{footnote}}

\setcounter{page}{1}

\section{Introduction}

Black holes are among the most fundamental and, at the same time, most fascinating
solutions of the general theory of relativity (GR). The different types of black holes
predicted by GR have all been determined and classified according to their physical
properties (mass, charge, angular momentum), and uniqueness theorems have been
formulated (see, for example, \cite{MTW, Carter}). The emergence of theories
\cite{ADD1, ADD2, ADD3, RS1, RS2} based on the early concept of brane \cite{Misha, Akama} and postulating
the existence of extra spacelike dimensions in nature has completely changed the 
landscape. Now, the higher-dimensional analogues of black holes cannot easily
be classified or proven to be unique---moreover, they are supplemented by a large number
of black objects such as black strings, black branes, black rings, or black saturns
\cite{Emparan-review}. 

In the limit where the self-energy of the brane is much smaller than the black-hole
mass and the symmetries of the four-dimensional solutions may be extended to
the higher-dimensional spacetime \cite{ADD1, ADD2, ADD3}, analytical forms of higher-dimensional
black holes, either spherically symmetric or rotating, are easy to derive---in fact,
they were derived long ago in \cite{Tangherlini, MP}. If, however, the brane self-energy
is not negligible and contributes toward a particular profile of gravity along
the extra dimensions \cite{RS1, RS2}, the analytic derivation of higher-dimensional black
holes is extremely difficult. A first attempt \cite{CHR}, employing a straightforward
ansatz of a Schwarzschild line element embedded into a warped extra dimension,
has failed to lead to a five-dimensional black hole and has instead led to a black
string, a five-dimensional solution with a horizon and a singularity at every point
of the extra dimension. It was subsequently shown that these black strings are
unstable under linear gravitational perturbations \cite{GL, RuthGL}; therefore these
unphysical objects are unlikely to survive in the context of a fundamental
gravitational theory. 

Despite decades of efforts, the quest for the more physically acceptable solutions
of regular, localized-close-to-our-brane black holes has failed to lead to an
analytical, nonapproximate form (see Refs. \cite{review1, review2, review3, review4, review5, review6, review7, tidal, Papanto, KT, KOT, CasadioNew, Frolov, Karasik, GGI, CGKM, Ovalle1, Ovalle2, Ovalle3, Ovalle4, Harko, daRocha1, daRocha2, Nakas}
for an impartial list of works)---however, such solutions were successfully derived in 
lower-dimensional gravitational models \cite{EHM1, EHM2, AS, Cuadros}. Numerical solutions also
emerged that described either small \cite{KTN, Kudoh, TT} or large black holes
\cite{Kleihaus, FW, Page1, Page2, Andrianov} in braneworld models. In an effort to derive the
long-sought analytical black-hole solutions, in \cite{KPZ, KPP2} the previously proposed
idea \cite{KT}, of adding a nontrivial profile along the extra dimension to the black-hole
mass function in the  original line element used in \cite{CHR}, was extended to include
also a dependence on the time and radial coordinate; in this way, the rather restricted
Schwarzschild-type of brane background was extended to include additional terms
[of an (anti)--de Sitter or Reissner-Nordstrom type] and to allow also for nonstatic
configurations. A large number of bulk scalar field theories were then investigated;
however, no viable solutions that could sustain the line element of a five-dimensional,
regular, localized-close-to-our-brane black hole was found.

In contrast, the analyses performed in \cite{KPZ, KPP2} have hinted toward the existence
of solutions that were not characterized by the desired nontrivial profile of the mass function
in terms of the extra coordinate---these solutions could not therefore be localized black holes
but rather novel black-string solutions. As a result, in this work, we focus on the careful
investigation of the existence of these latter types of solutions in the context of a theory
with a scalar field nonminimally coupled to gravity, and on the study of their physical
properties. We demonstrate that, for very natural, simple choices of the coupling function
between the scalar field and the five-dimensional scalar curvature, novel black-string
solutions may indeed be found with rather interesting and provocative characteristics.
Given the fact that the same theory has resisted in giving legitimate black-hole solutions,
even for a wider number of choices of the coupling function \cite{KPZ, KPP2}, our present
results add new ``fuel" to the long dispute around the question of why braneworld models
lead quite easily to black-string solutions but not to localized black holes
\cite{Fitzpatrick, Zegers, Heydari, Dai, Bruni, Dadhich, Kofinas, Tanaka, EFK, EGK, Yoshino}. Indeed, higher-dimensional gravitational theories
often allow for the emergence of uniform or nonuniform black-string solutions
\cite{Gubser, Wiseman, Kudoh2, Sorkin1, Sorkin2, Kleihaus2, Headrick, Figueras, Kalisch, Emparan2}.

In our analysis, we will retain the ``Vaidya form" of the brane line element used
also in \cite{KOT, KPZ, KPP2}, since this form was shown not to lead to additional
spacetime singularities in the bulk. As we are interested in finding static
black-string solutions, here we abandon the dependence of the mass function on
the time and extra-dimension coordinates, and allow for a general, radially dependent
form $m(r)$. Our field equations will be straightforwardly integrated to determine
the form of the mass function that is found to correspond to a Schwarzschild--anti--de
Sitter background. We will consider two simple forms of the coupling function, namely a
linear form and a quadratic one in terms of the scalar field. In  both cases, we solve the
set of field equations and derive the scalar-field configurations and the physical properties
of the model. 

A common characteristic of the solutions derived in the two cases is the negative sign
of the coupling function in front of the five-dimensional scalar curvature, either over
the entire bulk (for a quadratic dependence) or at distances larger than a specific value
(for a linear dependence). This clearly leads to the ``wrong sign" for gravity; however,
as we will see, it is this negative sign that effectively creates an anti--de Sitter spacetime 
and supports a Randall-Sundrum warp factor even in the absence of a negative bulk
cosmological constant. In the case of the linear coupling function, the antigravitating
regime arises away from our brane; this regime is pushed farther away the larger the
warping coefficient and the smaller the cosmological constant on our brane are. In fact,
for particular values of the parameters of the model, the theory resembles an ordinary,
minimally coupled scalar-tensor theory with normal gravity and a Randall-Sundrum
warp factor. 

Although the original objective of our analysis was to investigate the existence of novel
black-string solutions in the context of a nonminimally coupled scalar-tensor theory,
our solutions, in the limit of vanishing black-hole mass on the brane, reduce to
maximally symmetric braneworld solutions that are regular over the entire bulk
apart from the (anti)--de Sitter (AdS) boundary.\footnote{Braneworld solutions with a Minkowski spacetime
on our brane were also studied in the context of a nonminimally coupled scalar-tensor
theory in \cite{Farakos1, Bogdanos1, Farakos2, Farakos3}.} In this limit,
the gravitational background on our
brane is a pure AdS spacetime. In fact, we demonstrate that for the
physically motivated case of a positive-cosmological constant on our brane, the
emergence of an antigravitating regime in the bulk is unavoidable.

Our paper has the following outline: in Sec. 2, we present the field equations and
spacetime background. In Sec. 3, we study in detail the case of a linear coupling
function and determine the complete bulk solution, its physical properties, as well
as the effective theory on the brane. A similar analysis is performed in Sec. 4
for the case of a quadratic coupling function. In Sec. 5, we present a mathematical
argument that underlines the connection between the emergence of an antigravitating
regime in the bulk and the positive sign of the cosmological constant on our brane.
We finally present our conclusions in Sec. 6.


\section{The Theoretical Framework}

In this work, we focus on the following class of five-dimensional
gravitational theories with action functional
\beq
\label{action}
S_B=\int d^4x\int dy \,\sqrt{-g^{(5)}}\left[\frac{f(\Phi)}{2\kappa_5^2}R
-\Lambda_5-\frac{1}{2}\,\pa_L\Phi\,\pa^L\Phi-V_B(\Phi)\right].
\eeq
The above theory contains the five-dimensional scalar curvature $R$, a bulk
cosmological constant $\Lambda_5$, and a five-dimensional scalar field $\Phi$
with a self-interacting potential $V_B(\Phi)$ and a
nonminimal coupling to $R$ via the general coupling function $f(\Phi)$.
Also, $g^{(5)}_{MN}$ is the metric tensor of the
five-dimensional spacetime, and $\kappa_5^2=8\pi G_5$ is defined in terms of
the five-dimensional gravitational constant $G_5$. At a particular point along
the fifth spatial dimension, whose coordinate we denote by $y$, a 3-brane
is introduced---without loss of generality, we assume that this takes place
at $y=0$. Then, the above bulk action must be supplemented by the brane one
\beq
\label{13}
S_{br}=\int d^4x\sqrt{-g^{(br)}}(\lagr_{br}-\sigma)=
-\int d^4x\int dy\sqrt{-g^{(br)}}\,[V_b(\Phi)+\sigma]\,\delta(y)\,.
\eeq
Here, $\lagr_{br}$ is related to the matter/field content of the brane and has been
chosen to consist of an interaction term $V_b(\Phi)$ of the bulk scalar field with
the brane. Also, $\sigma$
is the brane self-energy, and $g^{(br)}_{\mu\nu}=g_{\mu\nu}^{(5)}(x^\lam,y=0)$ is
the induced-on-the-brane metric tensor. Note that, throughout our work, capital
letters $M,N,L,...$ will denote five-dimensional indices while lowercase letters
$\mu,\nu,\lambda,...$ will be used for four-dimensional indices.

\par The variation of the complete action $S=S_B+S_{br}$ with respect to the
metric-tensor components $g^{(5)}_{MN}$ yields the gravitational field equations
that have the form
\eq$\label{14}
f(\Phi)\,G_{MN}\sqrt{-g^{(5)}}=\kappa_5^2\left[(T^{(\Phi)}_{MN}-g_{MN}\Lambda_5)
\sqrt{-g^{(5)}}-[V_b(\Phi)+\sigma]\,g^{(br)}_{\mu\nu} \delta^\mu_M\delta^\nu_N\delta(y)\sqrt{-g^{(br)}}\right],$
where
\eq$\label{15}
T^{(\Phi)}_{MN}=\pa_M\Phi\,\pa_N\Phi+g_{MN}\left[-\frac{\pa_L\Phi\pa^L\Phi}{2}-V_B(\Phi)\right]
+\frac{1}{\kappa_5^2}\left[\nabla_M\nabla_Nf(\Phi)-g_{MN}\Box f(\Phi)\right].$
On the other hand, the variation of the action with respect to $\Phi$ leads to the
scalar-field equation
\beq
-\frac{1}{\sqrt{-g^{(5)}}}\,\pa_M\left(\sqrt{-\gfv}g^{MN}\pa_N\Phi\right)=
\frac{\pa_\Phi f}{2 \kappa^2_5} R-\pa_\Phi V_B 
-\frac{\sqrt{-g^{(br)}}}{\sqrt{-\gfv}}\,\partial_\Phi V_b\,\delta(y)\,\,.
\label{phi-eq-0}
\eeq

We will also assume that the five-dimensional gravitational background is given by
the expression
\eq$\label{metric}
ds^2=e^{2A(y)}\left\{-\left[1-\frac{2m(r)}{r}\right]dv^2+2dvdr+r^2(d\theta^2+\sin^2\theta d\varphi^2)\right\}+dy^2\,.$
The above line element is characterized by the presence of the warp factor
$e^{A(y)}$ that multiplies a four-dimensional background. For $m(r)=M$, this
four-dimensional line element is just the Vaidya transformation of the Schwarzschild
solution describing a black hole with mass $M$, and it leads to the same black-string
solutions found in \cite{CHR}. A generalized Vaidya form, where $m$ is not a
constant but a function of the coordinates, was used in a number
of works \cite{KOT, KPZ, KPP2} in an effort to find regular, localized black-hole
solutions. The motivation for the use of the Vaidya form of the four-dimensional
line element, instead of the usual Schwarzschild one, was provided
in \cite{KT, KOT}; in these, it was demonstrated that four-dimensional line elements
with horizons, such as the Schwarzschild one, when embedded in five-dimensional
spacetimes, transform their coordinate singularities at the horizons to true
spacetime ones \cite{KT}. In order to avoid this, in \cite{KOT}
the four-dimensional Schwarzschild line element was first transformed to its Vaidya
form and then embedded in the warped fifth dimension; in that case, no new
bulk singularities emerged. 

Although the desired black-hole solutions have not yet been analytically found in
braneworld models, the emergence of black-string solutions is more easily realized
\cite{CHR, Gubser, Wiseman, Kudoh2, Sorkin1, Sorkin2, Kleihaus2, Headrick, Figueras, Kalisch, Emparan2}. Indeed, in the context of the theory
(\ref{action}), hints for the
existence of novel black-string solutions described by the line element (\ref{metric})
were given in \cite{KPP2}. Therefore,  here we turn our attention to this question;
we will keep the general $r$-dependence of the mass function, i.e. $m=m(r)$, as
shown in Eq. (\ref{metric}), in order to
allow our brane metric background to deviate from the Schwarzschild form. Such a
modification may allow for terms proportional to an effective cosmological constant
or for terms of various forms associated with tidal charges to emerge. As the
explicit forms of the curvature invariant quantities for the line element (\ref{metric})
(given in Appendix \ref{App-Invar}) show, such a solution, if indeed supported by
the theory (\ref{action}), would describe a black-string solution with only the
black-hole singularity extended over the fifth dimension and no other singularity
present.

For the line element (\ref{metric}), one may easily see that the relation
$\sqrt{-g^{(5)}}=\sqrt{-g^{(4)}}$ holds, and the gravitational equations
are then simplified to 
\eq$\label{grav_new}
{f}(\Phi)\,G^M{}_N=T^{(\Phi)M}{}_N-\del^M{}_N\Lambda_5-[V_b(\Phi)+\sigma]\, g_{\mu\nu} g^{ML}\del^\mu_L\del^\nu_N\del(y),$
with
\beq\label{Tmn_new}
T^{(\Phi)M}{}_N=\pa^M\Phi\pa_N\Phi+\nabla^M\nabla_N {f}+\del^M{}_N(\lagr_\Phi-\Box {f}).
\eeq
In the above, we have defined
\beq\label{Lagr}\lagr_{\Phi}=-\frac{1}{2}\,\pa_L\Phi\,\pa^L\Phi-V_B(\Phi).
\eeq
In addition, for simplicity, we have absorbed the gravitational constant
$\kappa_5^2$ in the expression of the general coupling function $f(\Phi)$,
and omitted the superscripts
$^{(5)}$ and $^{(4)}$ from the bulk and brane metric tensors $g_{MN}$ and
$g_{\mu\nu}$, respectively. In fact, we will now focus on the gravitational
equations in the bulk and thus altogether remove the brane term proportional
to $\delta(y)$ from Eq. (\ref{grav_new})---when the junction conditions are
studied, this term will be reinstated.

The nonvanishing components of the Einstein tensor $G^M{}_N$ for the
background (\ref{metric}) are listed below:
\bea
&~&G^0{}_0=G^1{}_1=6A'^2+3A''-\frac{2e^{-2A}\pa_rm}{r^2},\nonumber \\[1mm] 
&~&G^2{}_2=G^3{}_3=6A'^2+3A''-\frac{e^{-2A}\pa_r^2m}{r}, \label{GMN} \\[2mm] 
&~&G^4{}_4=6A'^2-\frac{e^{-2A}\left(2\pa_rm+r\pa_r^2m\right)}{r^2}, \nonumber
\eea
where a prime ($'$) denotes the derivative with respect to the $y$-coordinate.
We will also assume that the bulk scalar field depends only on the coordinate
along the fifth dimension, i.e. $\Phi=\Phi(y)$. Then, the nonvanishing mixed
components of the energy-momentum tensor $T^{(\Phi)M}{}_N$ take in turn the form 
\begin{gather}
T^{(\Phi)0}{}_0=T^{(\Phi)1}{}_1=T^{(\Phi)2}{}_2=T^{(\Phi)3}{}_3=
A' \Phi'\,\pa_\Phi f+\lagr_\Phi-\Box f, \nonumber \\[2mm]
T^{(\Phi)4}{}_4=(1+\pa_\Phi^2 f)\Phi'^2+\Phi''\,\pa_\Phi f+\lagr_\Phi-\Box f,
\label{TMN-mixed}
\end{gather}
where, under the aforementioned assumptions, the quantities $\lagr_\Phi$ and $\Box f$ have
the explicit forms
\beq
\label{Lagr_new}
\lagr_\Phi=-\frac{1}{2}\,\Phi'^2-V_B(\Phi),
\eeq
and
\beq
\label{Box_f}
\Box f=4A' \Phi'\,\pa_\Phi f+\Phi'^2\,\pa_\Phi^2 f+\Phi''\,\pa_\Phi f.
\eeq

The gravitational field equations may now easily follow by substituting the components of
$G^M{}_N$ and $T^M{}_N$, listed in Eqs. (\ref{GMN}) and (\ref{TMN-mixed}), respectively,
in Eq. (\ref{grav_new}) evaluated in the bulk. We thus obtain three equations from the
$(^0{}_0)$, $(^2{}_2)$, and $(^4{}_4)$ components. Subtracting the $(^0{}_0)$ and $(^2{}_2)$
equations as well as the $(^0{}_0)$ and $(^4{}_4)$ equations, we arrive at two simpler 
ones that, together with the $(^0{}_0)$ component, form the following system
\beq
\label{eq-mass}
r\,\pa_r^2m-2\pa_rm=0\,,
\eeq
\beq \label{grav-1}
f\left(3A''+e^{-2A}\frac{\pa_r^2m}{r}\right)=\pa_\Phi f \left(A'\Phi'-\Phi''\right)
-(1+\pa_\Phi^2 f)\Phi'^2\,,
\eeq
\beq
\label{grav-2}
f\left(6A'^2+3A''-\frac{2e^{-2A}\pa_rm}{r^2}\right)=A'\Phi'\,\pa_\Phi f+
\lagr_\Phi -\Box f-\Lambda_5\,.
\eeq
The above gravitational equations are supplemented by the scalar-field equation of motion
(\ref{phi-eq-0}) that has the explicit form
\beq \label{phi-eq}
\Phi'' + 4A' \Phi' =\pa_\Phi f \left(10A'^2+4A''-e^{-2A}\frac{2\pa_rm+
r\,\pa_r^2m}{r^2}\right) +\pa_\Phi V_B\,.
\eeq

Equation (\ref{eq-mass}) can easily be integrated to yield the general form of the allowed
mass function, and this is
\beq
m(r)=M+ \Lambda r^3/6\,, \label{mass-sol}
\eeq
where $M$ and $\Lambda$ are arbitrary integration constants the physical interpretation
of which will be studied later (the coefficient 6 has been introduced for later convenience).
The above solution may now be used in order to simplify the form of the remaining
three equations (\ref{grav-1})--(\ref{phi-eq}). However, not all of them are independent:
as we explicitly demonstrate in Appendix \ref{App-Indep}, an appropriate manipulation
and rearrangement of the gravitational equations (\ref{grav-1}) and (\ref{grav-2})
lead to the same result following also from a similar manipulation of the scalar-field
equation (\ref{phi-eq}). Indeed, in a fully determined theory, i.e. with given $f(\Phi)$
and $V_B(\Phi)$, we would only need three independent equations out of the existing four
to find the two unknown metric functions $m(r)$ and $A(y)$ and the scalar field $\Phi(y)$. 
Therefore, henceforth, we will altogether ignore Eq. (\ref{phi-eq}) in our analysis and
retain Eqs. (\ref{grav-1}) and (\ref{grav-2}). We will then adopt the following approach:
we will assume the well-known form \cite{RS1, RS2} $A(y)=-k |y|$, with $k$ a positive constant,
for the warp factor of the
five-dimensional line element in order to ensure the localization of gravity near the brane; for
a chosen coupling function $f(\Phi)$, we will then determine the scalar-field configuration
by solving Eq. (\ref{grav-1}); finally, Eq. (\ref{grav-2}) will determine the form of the
potential $V_B(\Phi)$ that needs to be introduced to support the solution.

In the following sections, we present two simple choices for the coupling
function $f(\Phi)$, a linear one and a quadratic one; for each one, we determine the
corresponding solution for the scalar field and form of the potential and discuss
their physical characteristics.


\section{The Case of Linear Coupling Function}
\label{Linear}

\subsection{The bulk solution}
\label{Linear_1}

We will first consider the case where the coupling function is of the general linear
form, $f(\Phi)=a \Phi +b$, where $a$ and $b$ are constants. Employing this
together with the form of the mass function (\ref{mass-sol}) and the
exponentially decreasing warp factor $e^{A(y)}=e^{-ky}$ (assuming the usual ${\bf Z}_2$
symmetry in the bulk under the change $y \rightarrow -y$, we henceforth focus on the
positive $y$-regime), Eq. (\ref{grav-1}) takes the form
\beq
(a \Phi +b)\,\Lambda e^{2ky}=-a\,(k\,\Phi' + \Phi'') -\Phi'^2\,.
\label{grav-1-linear}
\eeq
In order to solve the above, we set $\Phi(y)=\Phi_0\,e^{g(y)}$. Substituting in 
Eq. (\ref{grav-1-linear}) and rearranging, we obtain
\beq
a\Lambda\Phi_{0}\,e^{2ky + g(y)} + b\Lambda\,e^{2ky} =
- a (kg'+ g'' + g'^{2})\Phi_{0}\,e^{g(y)}  -g'^{2}\Phi_{0}^{2}\,e^{2g(y)}\,,
\label{vasiki2}
\eeq
where a prime in $g$ denotes, as before, the derivative with respect to $y$. The above
leads to a nontrivial solution only if $g(y)=2ky$. In that case, the following
constraints should also hold:
\bea
a = -\frac{4k^2}{\Lambda}\,\Phi_{0}\,, \qquad 
b = \frac{24k^4}{\Lambda^2}\,\Phi_{0}^2\,. \label{consts-linear}
\eea
The coefficient $b$ is clearly positive definite; however the sign of the
coefficient $a$ depends on those of $\Phi_0$ and $\Lambda$. 

Let us examine the type of solution that we have derived. Employing the
form of the mass function (\ref{mass-sol}) and the general expressions for
the five-dimensional curvature invariants given in Appendix \ref{App-Invar}, the
latter quantities are found to have the form
\bea R &=& -20k^{2} + 4 \Lambda e^{2ky} \nonumber\\[2mm]
R_{MN}R^{MN} &=& 80 k^{4} - 32k^{2}\Lambda e^{2ky} + 4\Lambda^{2} e^{4ky}\,,
\label{invar}\\[1mm]
R_{MNKL}R^{MNKL} &=& 40k^{4} - 16k^{2}\Lambda e^{2ky} + \frac{8\Lambda^{2}e^{4ky}}{3}
+\frac{48 M^2 e^{4ky}}{r^6}\,. \nonumber 
\eea
For $M=\Lambda=0$, we recover the curvature invariants of the five-dimensional
AdS spacetime. For $\Lambda=0$ but $M \neq 0$, we obtain the black-string solution
of \cite{CHR}, with the black-hole singularity at $r=0$ extending over the entire
fifth dimension up to the AdS boundary at $ y \rightarrow \infty$. For $M=0$ but
$\Lambda \neq 0$, we find a solution that is everywhere regular apart from the
AdS boundary. Finally, for $M \neq 0$ and $\Lambda \neq 0$, we obtain again a
black-string solution with singular terms from both the black-hole and AdS boundary
appearing in the expressions of the curvature invariants. 

At this point, we should investigate the physical interpretation of the integration
constants $M$ and $\Lambda$ appearing in the expression (\ref{mass-sol}) of
the mass function $m(r)$. To this end, we set $y=0$ in the higher-dimensional
line element (\ref{metric}), and the projected-on-the-brane four-dimensional
background then reads
\eq$\label{metric-4D}
ds^2=-\left(1-\frac{2M}{r} -\frac{\Lambda r^2}{3}\right) dv^2+2dvdr+r^2(d\theta^2+
\sin^2\theta d\varphi^2)\,.$
The above looks like a generalization of the Vaidya form of the Schwarzschild line element
in the presence of a cosmological constant. In order to convince ourselves, we apply a
general coordinate transformation $v = h(t,r)$, where $h(t,r)$ will be defined shortly.
Then, Eq. (\ref{metric-4D}) assumes the standard, diagonal form
\eq$\label{metric-SdS}
ds^2=-\left(1-\frac{2M}{r} -\frac{\Lambda r^2}{3}\right) dt^2+
\left(1-\frac{2M}{r} -\frac{\Lambda r^2}{3}\right)^{-1} dr^2+r^2(d\theta^2+
\sin^2\theta d\varphi^2)$
provided that $h(t,r)=t +g(r)$ and $g(r)$ satisfies the following condition:
\beq
\frac{dg(r)}{dr}= \left(1-\frac{2M}{r} -\frac{\Lambda r^2}{3}\right)^{-1}.
\eeq
The details of the transformation as well as the explicit form of the function $g(r)$,
which is not of importance in the present analysis, can be found in Appendix
\ref{App-Vaidya}. According to Eq. (\ref{metric-SdS}),
the gravitational background on the brane is Schwarzschild--(anti)--de Sitter with $M$ being
the mass of the black hole and $\Lambda=\kappa_4^2 \Lambda_4$, where $\Lambda_4$
is the cosmological constant on the brane. 

It is of particular interest to study the profile of the nonminimal coupling
function $f(\Phi)$ along the extra dimension: using the solution for the scalar
field found above, we obtain
\beq
f(y) = a\,\Phi(y) +b= \frac{4k^2\Phi_{0}^2}{\Lambda^2}\,\left(-\Lambda e^{2ky}
+6k^2 \right)\,. \label{f-linear}
\eeq
For $\Lambda<0$, i.e. for a negative cosmological constant on the brane, the above expression
is everywhere positive definite and gravity remains attractive over the whole bulk. However, for
$\Lambda>0$, we find that
\beq
\left\{ \begin{tabular}{ll} $f(y)>0$\,, & $y<\frac{\ln(6k^2/\Lambda)}{2k}$ \\[2mm]
$f(y)=0$\,, & $y=y_0 \equiv \frac{\ln(6k^2/\Lambda)}{2k}$ \\[2mm]
$f(y)<0$\,, & $y>\frac{\ln(6k^2/\Lambda)}{2k}$ \end{tabular}  \right \}.
\label{f_lin_Lpos}
\eeq
That is, close to the brane and up to a maximum distance of $y=y_0$ the function
$f(y)$ is positive and gravity acts as normal. However, at $y=y_0$, $f(y)$
vanishes, and gravity locally disappears, whereas,  for $y>y_0$, $f(y)$ turns
negative, and gravity acquires the wrong sign. We may therefore conclude that,
for a positive cosmological constant on the brane, gravity becomes repulsive in the
bulk at some finite distance away from the brane.  

\begin{figure}[t!]
    \centering
    \begin{subfigure}[b]{0.46\textwidth}
        \includegraphics[width=\textwidth]{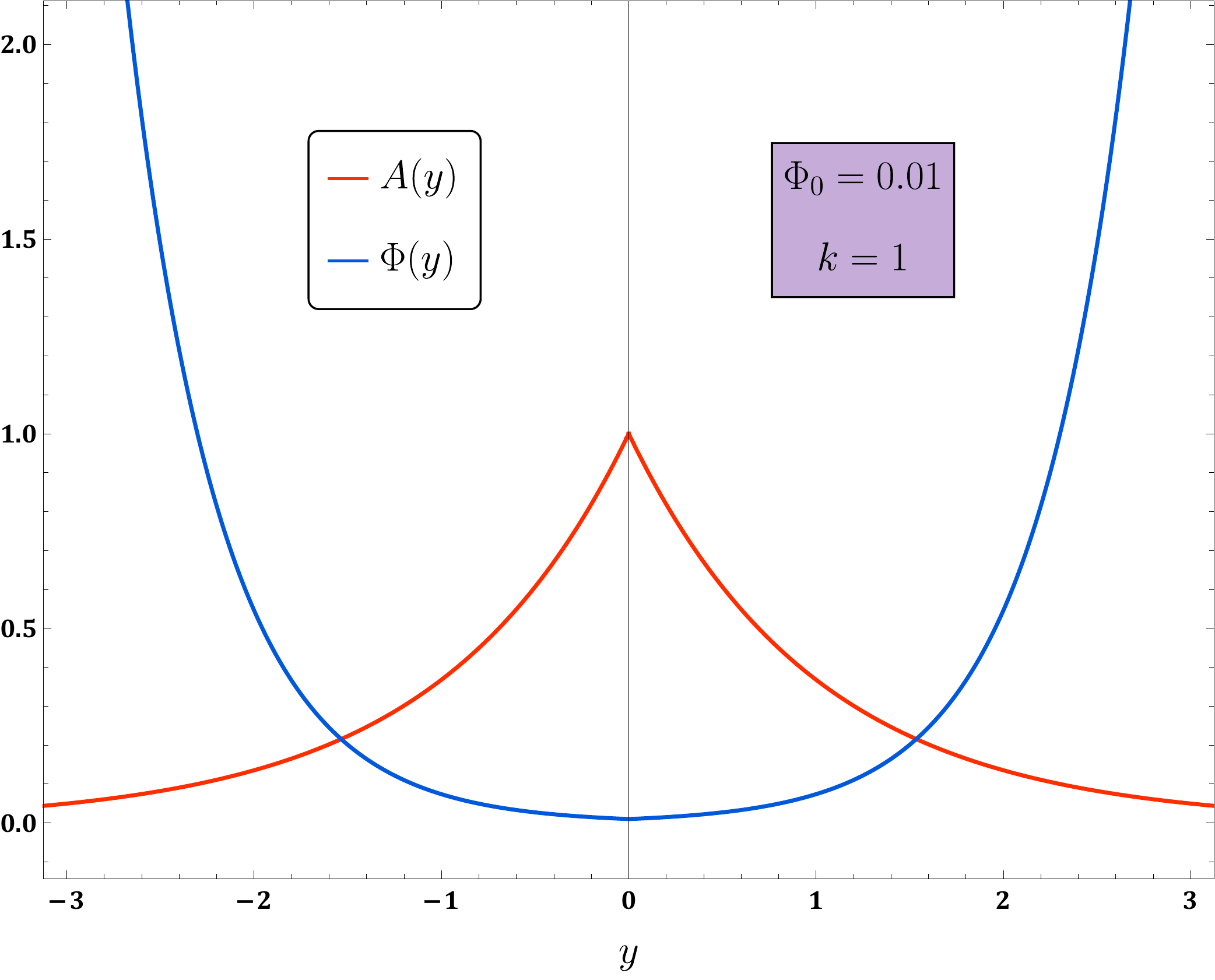}
        \caption{}
        \label{fig1a}
    \end{subfigure}
    ~ 
    \begin{subfigure}[b]{0.49\textwidth}
        \includegraphics[width=\textwidth]{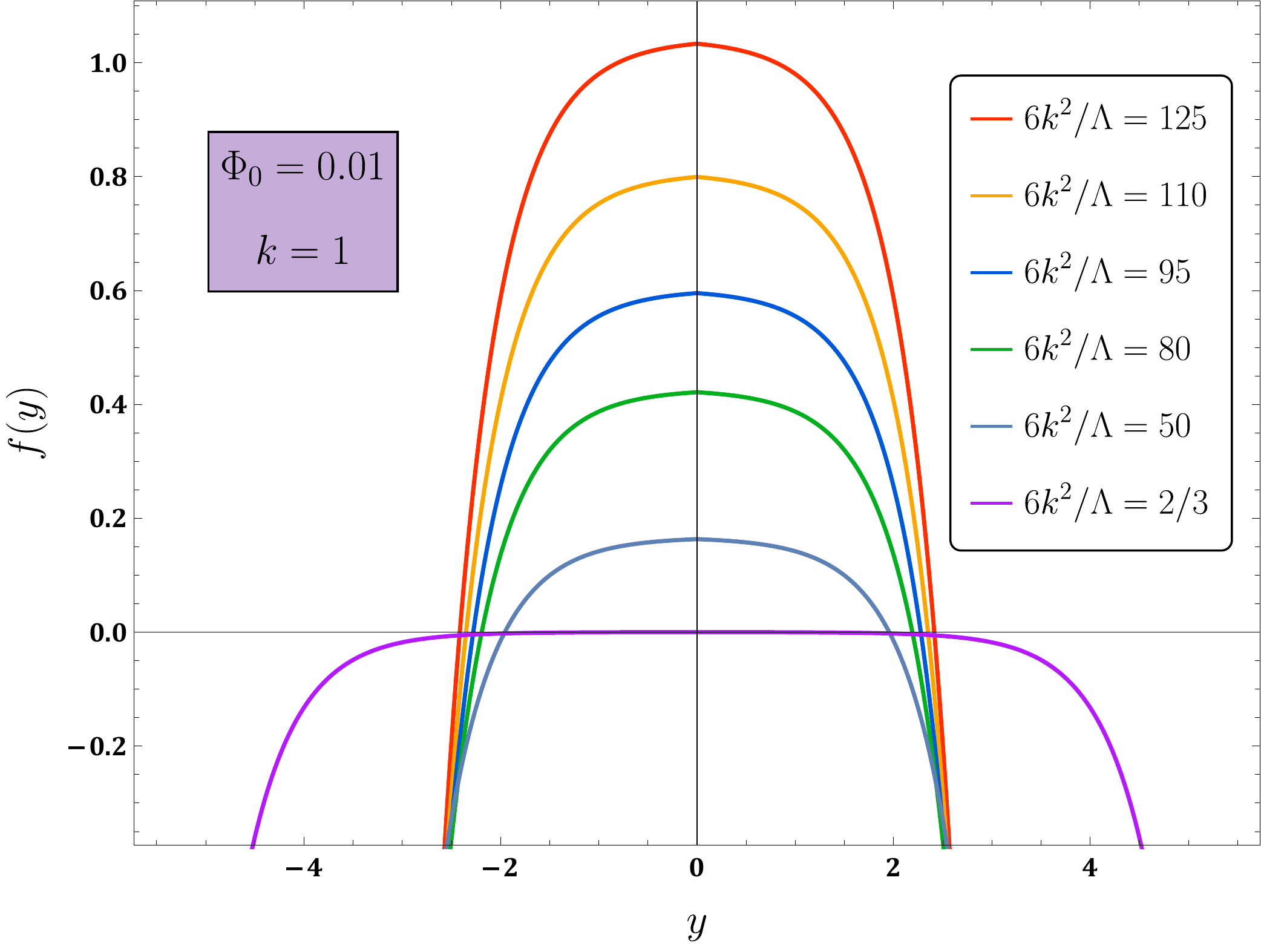}
        \caption{}
        \label{fig1b}
    \end{subfigure}
    ~ 
    \caption{(a) The warp factor $e^{-k |y|}$ and scalar field $\Phi(y)$, and (b)
     the coupling function $f(y)=a \Phi(y) +b$, in terms of the coordinate $y$,
   for $k=1$, $\Phi_0=0.01$, and $6k^2/\Lambda=2/3,50,80,95,110,125$ (from bottom to top).}
   \label{warp_Phi_f}
\end{figure}

In Fig. 1(a), we depict the form of the warp factor $e^{-k |y|}$
and the scalar field $\Phi(y)$ in terms of the coordinate $y$ along the fifth dimension,
for $k=1$ and $\Phi_0=0.01$. 
Although the former quantity exhibits the anticipated localization close to our brane,
the latter quantity increases away from the brane diverging at the boundary of
spacetime. The displayed, qualitative behavior of these two quantities is independent
of the particular values of the parameters. In contrast to this, the profile of the
coupling function  $f(y)$, given in Eq. (\ref{f-linear}), depends strongly on the value
of the dimensionless parameter $6k^2/\Lambda$: assuming that $\Lambda>0$ on our brane,
in Fig. 1(b) we display the form of $f(y)$, for $k=1$, $\Phi_0=0.01$, and
the values $6k^2/\Lambda=2/3,50,80,95,110,125$. For $6k^2/\Lambda <1$, the
function $f(y)$ does not have
a vanishing point and is always negative; for $6k^2/\Lambda >1$, a regime of positive
values for $f(y)$ appears close to our brane that tends to become larger as $k^2/\Lambda$
gradually increases. In other words, the smaller the cosmological constant
is on our brane, the farther away from our brane the antigravitating regime is
located. It is also interesting to note that the regime of positive values for the
function $f(y)$ around our brane is always characterized by a plateau, an area where the
value of the coupling function is almost constant; therefore, close to our brane,
gravity would not only act as normal but it would look as if the scalar curvature
$R$ does not have a coupling to the scalar field. In fact, for the particular value
of $6k^2/\Lambda=125$, the coupling function $f(\phi)$ around the brane is constant
and approximately equal to unity; thus the model mimics ordinary, five-dimensional
gravity with the difference that the bulk energy, which as we will see supports
the complete bulk-brane solution, originates, in fact, from the scalar field.

In order to complete the analysis, we need to determine the potential of the scalar
field $V_B(\Phi)$. Substituting the forms of the functions $m(r)$, $A(y)$, and $\Phi(y)$
in Eq. (\ref{grav-2}), we readily obtain
\beq
V_B(\Phi)=-\Lambda_5 -2 k^2 \left(\frac{72 k^4 \Phi_0^2}{\Lambda^2} -
\frac{20 k^2 \Phi_0}{\Lambda}\,\Phi +3 \Phi^2\right)\,.
\label{V_linear}
\eeq
Combining the above expression with the profile of the scalar field along the extra dimension,
$\Phi(y)=\Phi_0\,e^{2ky}$, we notice the following: at the location of the brane, at $y=0$,
the scalar potential reduces to a constant value, namely
\beq
V_B(y=0)=-\Lambda_5 -2 k^2 \Phi_0^2 \left(\frac{72 k^4}{\Lambda^2} -
\frac{20 k^2}{\Lambda} +3 \right).
\label{V_linear_y0}
\eeq
The quantity inside the brackets has no real roots and is thus always positive definite;
that makes the second term a negative-definite quantity for all values of the parameters
of the model. This means that, close to the brane, the scalar potential can mimic the
role of the negative cosmological constant---thus making $\Lambda_5$ redundant---and
support by itself an AdS spacetime in the bulk regime close to the brane. 

In Fig. 2(a), we depict the form of the scalar potential
found above, for the choice of parameters $6k^2/\Lambda=100$, $\Phi_0=0.01$, $k=1$, and for
$\Lambda_5=0$. The regime close to our brane where $V_B$ mimics the negative cosmological
constant is clearly present. As we move away from the brane, the scalar field starts
increasing: this leads first to the formation of a small barrier (i.e. a local extremum),
as a result of the competing roles of the linear and quadratic in $\Phi$ terms in
Eq. (\ref{V_linear}), and eventually to the divergence of $V$ toward minus infinity
at the boundary of spacetime. In the same plot, we depict the form of the coupling
function $f(y)$, for the same parameter values: this ensures us of the fact that the 
regime of the mimicking of ``negative cosmological constant" and the location of the
barrier lies well inside the normal gravitating regime. At the point where $f(y)$ 
vanishes and gravity disappears, $V(y)$ retains a moderate, finite value; allowing one,
however, to enter the antigravitating regime leads to arbitrary large negative values
of the scalar potential.

\begin{figure}[t!]
    \centering
    \begin{subfigure}[b]{0.47\textwidth}
        \includegraphics[width=\textwidth]{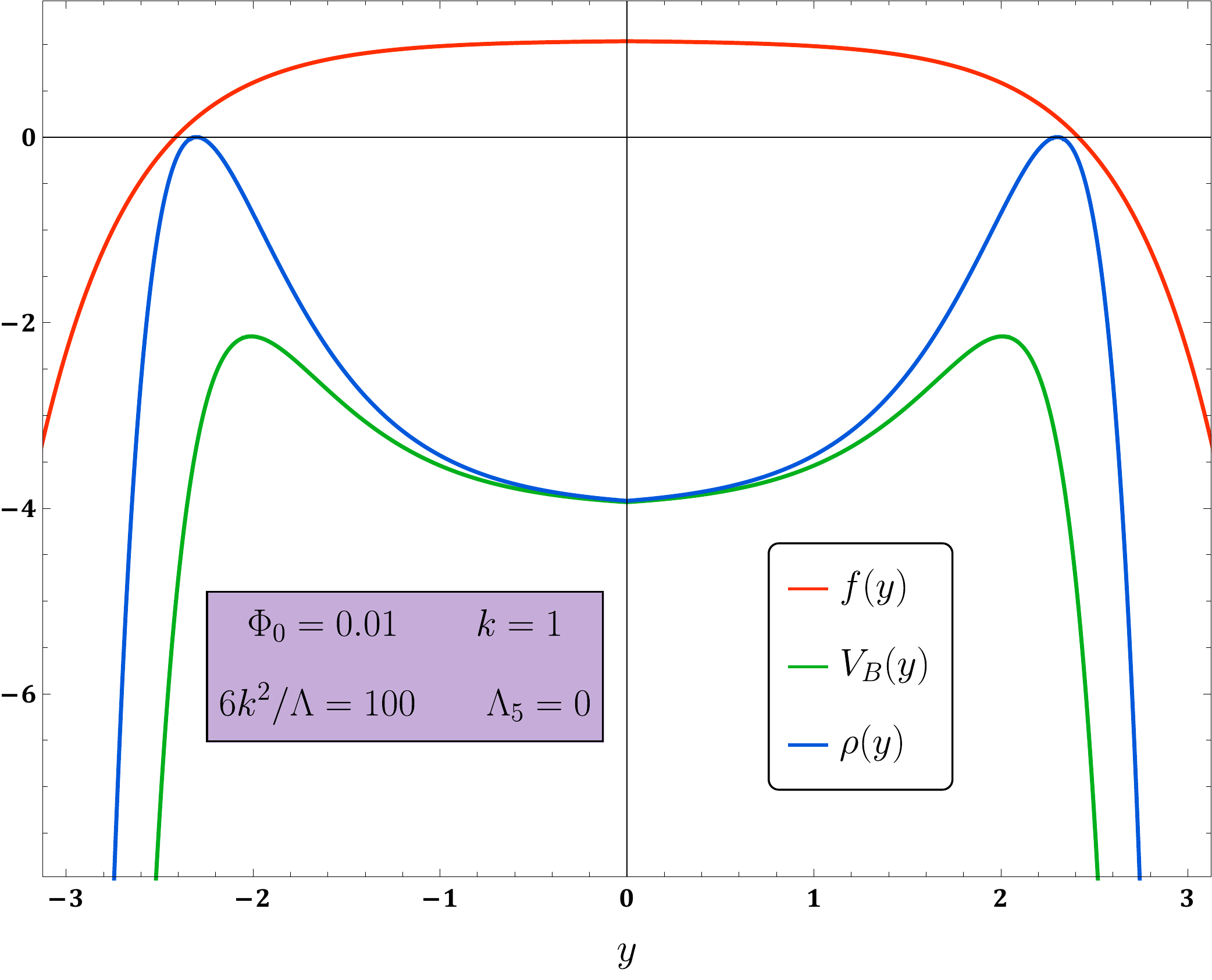}
        \caption{}
        \label{fig2a}
    \end{subfigure}
    \quad
    ~ 
    \begin{subfigure}[b]{0.47\textwidth}
        \includegraphics[width=\textwidth]{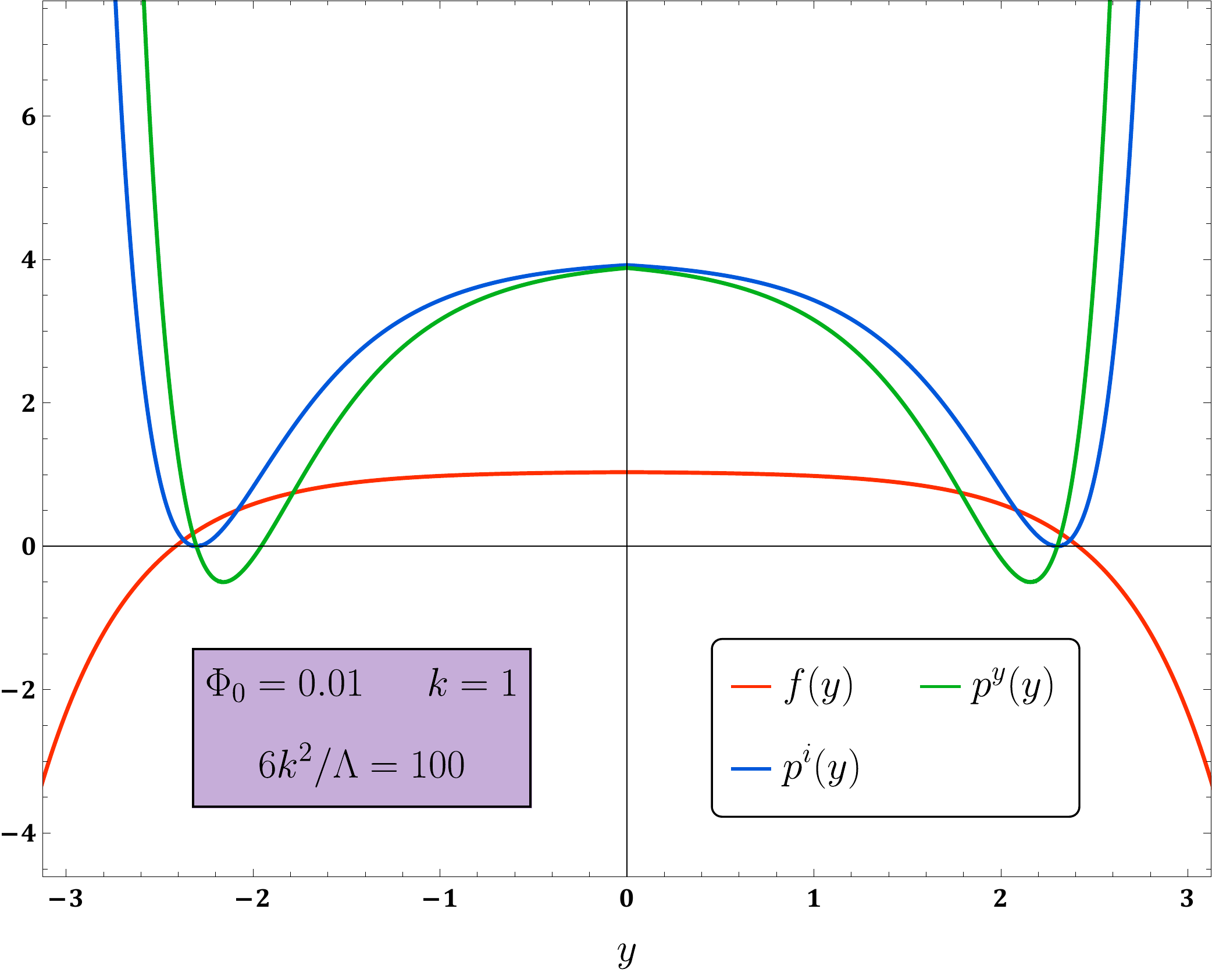}
        \caption{}
        \label{fig2b}
    \end{subfigure}
    ~ 
    \caption{(a) The scalar potential $V_B$ and energy density $\rho$ of the system, and
(b) the pressure components $p^y$ and $p^i$ in terms of the coordinate $y$
(from bottom to top in both plots), for $6k^2/\Lambda=100$, $\Phi_0=0.01$, $k=1$,
and $\Lambda_5=0$. We also display the coupling function $f$ with its characteristic plateau,
for easy comparison.}
   \label{V_Tmn_Linear}
\end{figure}

The components of the energy-momentum tensor of the theory may be computed employing
Eqs. (\ref{grav_new}) and (\ref{TMN-mixed}). Using also the relations $\rho=-T^0_0$,
$p^i=T^i_i$, and $p^y=T^y_y$, we find the explicit expressions
\beq
\rho=-p^i=\Lambda_5 -2a k^2\Phi + 2k^2 \Phi^2 + V_B(\Phi)
= - 4 k^2 \Phi_0^2\left(\frac{6k^2}{\Lambda}-e^{2ky}\right)^2\,,
\label{T00_linear}
\eeq
\beq
p^y=-\Lambda_5 + 8a k^2 \Phi + 2k^2 \Phi^2 -V_B(\Phi)
= 8 k^2 \Phi_0^2\left(\frac{18 k^4}{\Lambda^2}-\frac{9 k^2}{\Lambda}\,e^{2ky}
+e^{4ky}\right)\,.
\label{Tyy_linear}
\eeq
The behavior of the above quantities is also depicted in Fig. \ref{V_Tmn_Linear},
for the same values of parameters as in Fig. \ref{warp_Phi_f} to allow for an easy comparison.
As expected, close to the brane the profile of all components resembles that of a
negative cosmological constant. At an intermediate distance, the energy density $\rho$
reaches a local, maximum value, and, far away from the brane---inside the
antigravitating regime---it diverges to negative infinity. The pressure components
$p^y$ and $p^i$ exhibit the exact opposite behavior: starting from a constant
value near the brane, they dive toward a local minimum and, inside the antigravitating
regime, diverge to positive infinity. We readily observe that the total energy density
$\rho$ of the system remains negative throughout the bulk---this is also obvious from
its expression in Eq. (\ref{T00_linear}); however, this is due to a physical, scalar field
with a potential that turns out to be negative in order to create the local AdS spacetime
and support the decreasing warp factor. Close to the brane, that potential is analytic
and finite---should one wish to ban the diverging, antigravitating regime from the
bulk spacetime, a second brane could easily be introduced at a distance $y=L<y_0$.  
The necessity of introducing a second brane in the model will be discussed shortly.

\subsection{Junction conditions and effective theory}

Let us now address the issue of the junction conditions introduced in the
model due to the presence of the brane at $y=0$. The energy content of the brane
is given by the combination $\sigma + V_b(\Phi)$, and it creates a discontinuity
in the second derivatives of the warp factor and scalar field at the location of
the brane. We write $A''=\hat A'' + [A']\,\delta(y)$ and $\Phi''=\hat \Phi'' +
 [\Phi']\,\delta(y)$, where the hat quantities denote the distributional (i.e. regular)
parts of the second derivatives and $[\cdots]$ gives the discontinuities of the corresponding
first derivatives across the brane \cite{BDL}. Then, if we reintroduce the delta-function
terms in both the Einstein equation (\ref{grav-1}) and the scalar-field equation (\ref{phi-eq}),
and match the coefficients of the delta-function terms, we obtain the conditions
\bea
&3 f(\Phi)\,[A'] = -[\Phi']\,\partial_\Phi f- (\sigma + V_b)\,,& \\[2mm]
& [\Phi'] = 4 [A']\,\partial_\Phi f + \partial_\Phi V_b\,,&
\label{jun_linear}
\eea
where all quantities are evaluated at $y=0$.  Using the expressions for the warp factor
and the scalar field, as well as the symmetry in the bulk under the change
$y \rightarrow -y$, we find their explicit forms
\bea
\frac{8 k^2}{\Lambda}\,k \Phi_0^2 \left(1-\frac{18 k^2}{\Lambda}\right) &=&
-\sigma -V_b(\Phi_0)\,, \label{jun_linear_ex_1}\\[2mm]
4k\Phi_0\left(1-\frac{8k^2}{\Lambda}\right)&=&\partial_\Phi V_b\bigl|_{y=0}\,.
\label{jun_linear_ex_2}
\eea
According to the second junction condition, in the absence of an interaction term $V_b$
of the scalar field with the brane, we should have $k^2=\Lambda/8$. This result
determines the sign of the four-dimensional cosmological constant that must necessarily
be positive and relates its magnitude to the scale of warping in the bulk.
Moreover, the dimensionless quantity $k^2/\Lambda$, which determines the range
of the gravitating regime, should be exactly $1/8$: this value, being smaller than $1/6$,
does not allow for a normal gravity regime anywhere in the bulk, according to the
discussion above. The first of the conditions also leads to the result $\Phi_0^2=4\sigma/5k$;
for the case $k>0$, which ensures the decrease of the warp factor away from our brane,
the brane self-energy $\sigma$ comes out to be positive, too, and thus is physically
acceptable. 

As we showed above, the existence of a normal-gravity regime close to our brane
demands the presence of an interaction term $V_b$ of the scalar field with the brane.
Although the number of choices for $V_b$ is
in this case infinite, one may draw some general conclusions: if we assume again that
$k>0$ and that $k^2/\Lambda>1/6$, so that a positive $f(\Phi)$-regime exists around
our brane, then Eq. (\ref{jun_linear_ex_1}) still ensures that the total
energy content of our brane $\sigma + V_b(\Phi_0)$ is always positive. Assuming now,
as an indicative case, a linear form for the interaction term, too, i.e.
$V_b(\Phi)=\lambda_0\,\Phi$, where $\lambda_0$ is a coupling constant, we obtain the
conditions
\beq
\frac{8 k^2}{\Lambda}\,k \Phi_0^2 \left(1-\frac{18 k^2}{\Lambda}\right) =
-\sigma -\lambda_0 \Phi_0\,, \qquad
4k\Phi_0\left(1-\frac{8k^2}{\Lambda}\right)=\lambda_0\,.
\label{jun_linear_example}
\eeq
The above two conditions determine two out of the five parameters of the model:
$(k, \Lambda, \Phi_0, \lambda_0, \sigma)$. Considering the bulk scalar field and the
self-energy of the brane as the constituents of the model that support the complete
bulk-brane solution, the parameters related to them, namely the value of the field
on the brane $\Phi_0$, its coupling constant with the brane $\lambda_0$, and the brane
self-energy $\sigma$, may be naturally chosen as the true independent quantities of
the theory. On the other hand, the scale of the warping $k$ and the effective
cosmological constant on the brane $\Lambda$ are determined through the junction
conditions by the aforementioned three fundamental parameters. In this case, one
may easily see that, for $\lambda_0 \Phi_0>0$, we obtain $k^2/\Lambda<1/8$, which
allows for a bulk that is everywhere antigravitating, while, for $\lambda_0 \Phi_0<0$,
solutions with large values of $k^2/\Lambda$ may be obtained that have their
antigravitating regime pushed away from our brane.

We should finally address the issue of the effective theory on the brane. The negative sign
of the coupling function $f(\Phi)$ emerging at some distance from the brane as well
as the diverging behavior of the field $\Phi$ in the same region raise concerns
about the type of the effective theory that a four-dimensional observer would witness.
In order to answer this question, we need to derive the four-dimensional effective
action by integrating the five-dimensional one, given in Eq. (\ref{action}), over the
fifth coordinate $y$. Employing the first of Eqs. (\ref{invar}), we write
$R=-20k^2 +R^{(4)} e^{2ky}$, where $R^{(4)}=4 \Lambda$ is the scalar
curvature of the projected-on-the-brane line element (\ref{metric-4D}). Then, the
action takes the form
\beq
S=\int d^4 x\,dy\,\sqrt{-g^{(5)}}\left[\frac{f(\Phi)}{2}\,\left(e^{2ky} R^{(4)}-
20k^2\right) -\Lambda_5-\frac{1}{2}\,\Phi'^2 -V_B(\Phi)\right]. \label{action_eff}
\eeq
Using also that $\sqrt{-g^{(5)}}=e^{-4k|y|} \sqrt{-g^{(4)}}$, the four-dimensional,
effective gravitational constant would be given by the integral
\beq
\frac{1}{\kappa_4^2}\equiv 2 \int_{0}^{\infty} dy\, e^{-2 k y}\,f(\Phi)
=\frac{8k^2 \Phi_0^2}{\Lambda^2}\,\int_{0}^{\infty} dy\,\left(-\Lambda +
6k^2 e^{-2 k y}\right)\,.
\label{effG_1}
\eeq
In the above, we have substituted the form of the coupling function $f(\Phi)$ given in
Eq. (\ref{f-linear}). We observe that, although the second term inside the brackets will
lead to a finite result even for a noncompact fifth dimension---similar to the
Randall-Sundrum model, the first term will give
a divergent contribution. As a result, the presence of a second brane at a distance
$y=L$ is imperative for a well-defined effective theory. In that case, the upper limit
of the $y$-integral in Eq. (\ref{effG_1}) is replaced by $L$, and we obtain
\beq
\frac{M_{Pl}^2}{8\pi}=\frac{\Phi_0^2}{k}\,\frac{8 k^2}{\Lambda} \left[
\frac{3k^2}{\Lambda}\,(1-e^{-2 k L})-kL\right]\,.
\label{effG_2}
\eeq
Compared to the Randall-Sundrum model \cite{RS1, RS2}, the expression for the four-dimensional
gravity scale $M_{Pl}^2$ involves the quantity $\Phi_0^2$---with units $[M]^3$---and
the dimensionless parameter $k^2/\Lambda$ on its right-hand side. This signifies the fact
that, in the context of the theory (\ref{action}), the five-dimensional gravity scale $M_5^3$
may altogether be replaced by the coupling function $f(\Phi)$. If one chooses large values
for the $k^2/\Lambda$ parameter, then the value of the effective Planck scale may differ
from that of $\Phi_0$ by orders of magnitude. In fact, the smaller the cosmological constant
is on our brane, the more extended is the positive-value regime for $f(\Phi)$, as we
saw in the previous subsection, and the larger the difference between $M_{Pl}^2$
and $\Phi_0^2$. Equation (\ref{effG_2}) contains also a term linear in the interbrane
distance $L$, which was absent in the Randall-Sundrum case. Therefore, one should
take care that the inequality $kL <3k^2/\Lambda$ is always satisfied---however,
for small values of the four-dimensional cosmological constant on the brane, as argued
above, this constraint should easily be satisfied. 

The introduction of the second brane in order to ensure a finite effective theory
on our brane is supplemented by a second set of junction conditions at the location
$y=L$. A brane source term of the form $-[\hat \sigma + \hat V_b(\Phi)]\,\delta(y-L)$
should be introduced in the action, where $\hat \sigma$ and $\hat V_b(\Phi)$ are
the self-energy of the second brane and the interaction term of the scalar field
with that brane, respectively. We follow a similar procedure as at $y=0$ and
arrive at a set of junction conditions similar to those in Eq. (\ref{jun_linear})
but with all quantities evaluated at $y=L$. Their explicit form reads
\bea
\frac{8 k^2}{\Lambda}\,k \Phi_0^2 \left(\frac{18 k^2}{\Lambda}-e^{2kL}\right) &=&
-\hat\sigma -\hat V_b(\Phi)\bigl|_{y=L}\,, \label{jun_linear_ex_1_2}\\[2mm]
4k\Phi_0\left(\frac{8k^2}{\Lambda}-e^{2kL}\right)&=&\partial_\Phi \hat V_b\bigl|_{y=L}\,.
\label{jun_linear_ex_2_2}
\eea
The above set of conditions may be used to determine two more parameters of the
model: one may be the interbrane distance $L$ and the other a parameter 
associated with the interaction term $\hat V_b$. The self-energy of the second
brane $\hat \sigma$ as well as the functional form of $\hat V_b(\Phi)$ remain
completely arbitrary.

To complete the derivation of the effective theory on the brane, we finally
compute the effective cosmological constant---this may be used as a consistency
check of our results. The cosmological constant on the brane is given by the
integral of the remaining terms in Eq. (\ref{action_eff})---since $\Phi$ is
only $y$-dependent, no dynamical field will survive in the effective theory.
These terms will be supplemented by the source terms of the two branes as well as
the Gibbons-Hawking terms at the boundaries of spacetime \cite{Gibbons-terms}.
In total, we will have
\bea
-\Lambda_4&=&\int_{-L}^L dy\,e^{-4k|y|}\left[-10 k^2 f(\Phi) -\Lambda_5-
\frac{1}{2}\,\Phi'^2 -V_B(\Phi) +f(\Phi)(-4A'')|_{y=0}\right. \nonumber \\[2mm]
&& \hspace*{1cm}\left.+f(\Phi)(-4A'')|_{y=L} -[\sigma +V_b(\Phi)]\,\delta(y)-
[\hat \sigma + \hat V_b(\Phi)]\,\delta(y-L)\right].
\label{L-eff-1}
\eea
Employing the expressions for the coupling function and scalar potential, Eqs. (\ref{f-linear})
and (\ref{V_linear}), respectively, as well as the junction conditions (\ref{jun_linear_ex_1}) and
(\ref{jun_linear_ex_1_2}), and integrating over $y$, we finally obtain the result
\beq
\Lambda_4= 8k^2 \Phi_0^2\left[\frac{3k}{\Lambda} \left(1-e^{-2kL}) -L\right)\right] =
\frac{\Lambda}{\kappa_4^2}\,,
\label{L-eff-2}
\eeq
where we have used the expression for the effective gravitational scale $M_{Pl}^2/8\pi=1/\kappa_4^2$
given in Eq. (\ref{effG_2}).
As expected, the derivation of the effective theory has confirmed the interpretation of the
metric parameter $\Lambda$ as the product $\kappa_4^2 \Lambda_4$, that followed
also by comparing the projected-on-the-brane line element (\ref{metric-4D}) with
the standard Schwarzschild--de Sitter background.

We finally to note that the presence of the mass parameter $M$ has played no
role either in the profile of the scalar field and the energy-momentum tensor components
or in the derivation of the junction conditions and the effective theory on the brane.
Its presence creates a Schwarzschild--de Sitter background on the brane and an extended
singularity into the bulk leading to a five-dimensional black string stretching between the two
branes. If we set this parameter equal to zero, then the four-dimensional background on
the brane reduces to a pure de Sitter spacetime while the five-dimensional background
is free of singularities as long as $L<\infty$. For $L>y_0$, the bulk will also contain
an antigravitating regime (unavoidable for $\Lambda_4>0$, as we will see in the next
section).


\section{The Quadratic Case}

We now move to the case where the coupling function has a quadratic form, i.e.
$f(\Phi)=a\,\Phi^2$, where $a$ is again a constant. Employing, as in the previous
subsection, the form of the mass function (\ref{mass-sol}) and the warp factor
$e^{A(y)}=e^{-ky}$,  Eq. (\ref{grav-1}) now takes the form
\beq
a \Lambda e^{2ky} \Phi^2=-2a\,\Phi\,(k\,\Phi' + \Phi'') -(1+2a)\Phi'^2\,.
\label{grav-1-quad}
\eeq
Again, we set: $\Phi(y)=\Phi_0\,e^{g(y)}$, and the above equation is rewritten as
\beq
a \Lambda e^{2ky} =-2a\,(k g' + g'') -(1+4a)g'^2\,.
\label{grav-1-quad_2}
\eeq
The above calls for an exponential dependence for the function $g(y)$---we thus set
$g(y)=g_0\,e^{\lambda y}$, with $g_0$ and $\lambda$ constant coefficients, and write
the above equation as
\beq
a\Lambda\,e^{2ky} = - 2a g_0 \lambda\,(k+\lambda)\,e^{\lambda y}  
-(1+4a) g_0^2 \lambda^2\,e^{2\lambda y}\,.
\label{vasiki2_quad}
\eeq
There is again only one nontrivial solution that satisfies the aforementioned equation,
and this corresponds to the choice $\lambda=2k$. Then, the following constraints
should hold:
\bea
a = -\frac{1}{4}\,, \qquad \quad
g_0 = -\frac{\Lambda}{12 k^2}\,. \label{consts-quad}
\eea
The coefficient $a$ is negative definite, and therefore in this case gravity acts as a repulsive
force over the entire bulk. We should note here that an attempt to generalize the form of
the coupling function according to the ansatz $f(\Phi)=a\,\Phi^2 + b\,\Phi +c$, where
($a,b,c$) are constant coefficients, failed to lead to a consistent solution. Had such a
solution been possible, we could perhaps find regimes in the $y$-coordinate where gravity
would act as normal, hopefully close to our brane. Unfortunately such a solution has not
emerged, and therefore for a quadratic coupling function, the theory always leads to an
antigravitating bulk. This feature is strongly connected to the presence of the
cosmological constant on the brane---we will return to this point in the following
section.

Let us, however, investigate the remaining aspects of the model. The warp factor 
assumes the standard Randall-Sundrum form, i.e. $e^{A(y)}=e^{-k |y|}$, and is displayed
in Fig. 3(a). The profile of the scalar field depends on the sign of the
parameter $\Lambda$: using the second of the constraints (\ref{consts-quad}), we find that
\beq
\Phi(y)=\Phi_0\,\exp\left(-\frac{\Lambda}{12k^2}\,e^{2ky}\right)\,.
\label{Phi_quadr}
\eeq
The projected-on-the-brane line element is still given by Eq. (\ref{metric-4D});
thus, $\Lambda$ is again proportional to the brane cosmological constant. Then,
Eq. (\ref{Phi_quadr}) tells us that, for a positive cosmological constant on the brane,
the scalar field takes its maximum value $\Phi=\Phi_0 e^{-\Lambda/12k^2}$ at the
location of our brane, and decreases fast as we move away from the brane.
Therefore, the scalar field exhibits a localization around our brane similar to
that of the warp factor; in particular, for small values of the parameter $k^2/\Lambda$,
the profile of the scalar field exhibits a cusp at the location of the brane ($y=0$)
while, as $k^2/\Lambda$ increases, a plateau appears around the brane.
The coupling function, $f(\Phi)=a \Phi^2$,
assumes a similar profile by decreasing very fast, as $y$ increases; as a result,
the antigravitating regime associated with $f(y)$ is rather small. The 
profiles of the scalar field and coupling function, for $\Lambda>0$, are 
depicted in Figs. 3(a) and 3(b), respectively. 
On the other hand, for a negative cosmological constant on the brane, the scalar
field increases very fast away from the brane blowing up at the boundary of the
spacetime, and the same behavior is exhibited by the coupling function $f(\Phi)$. 
In what follows, we ignore this unattractive solution and explore further the more
interesting one with a positive cosmological constant on the brane.

\begin{figure}[t!]
    \centering
    \begin{subfigure}[b]{0.44\textwidth}
        \includegraphics[width=\textwidth]{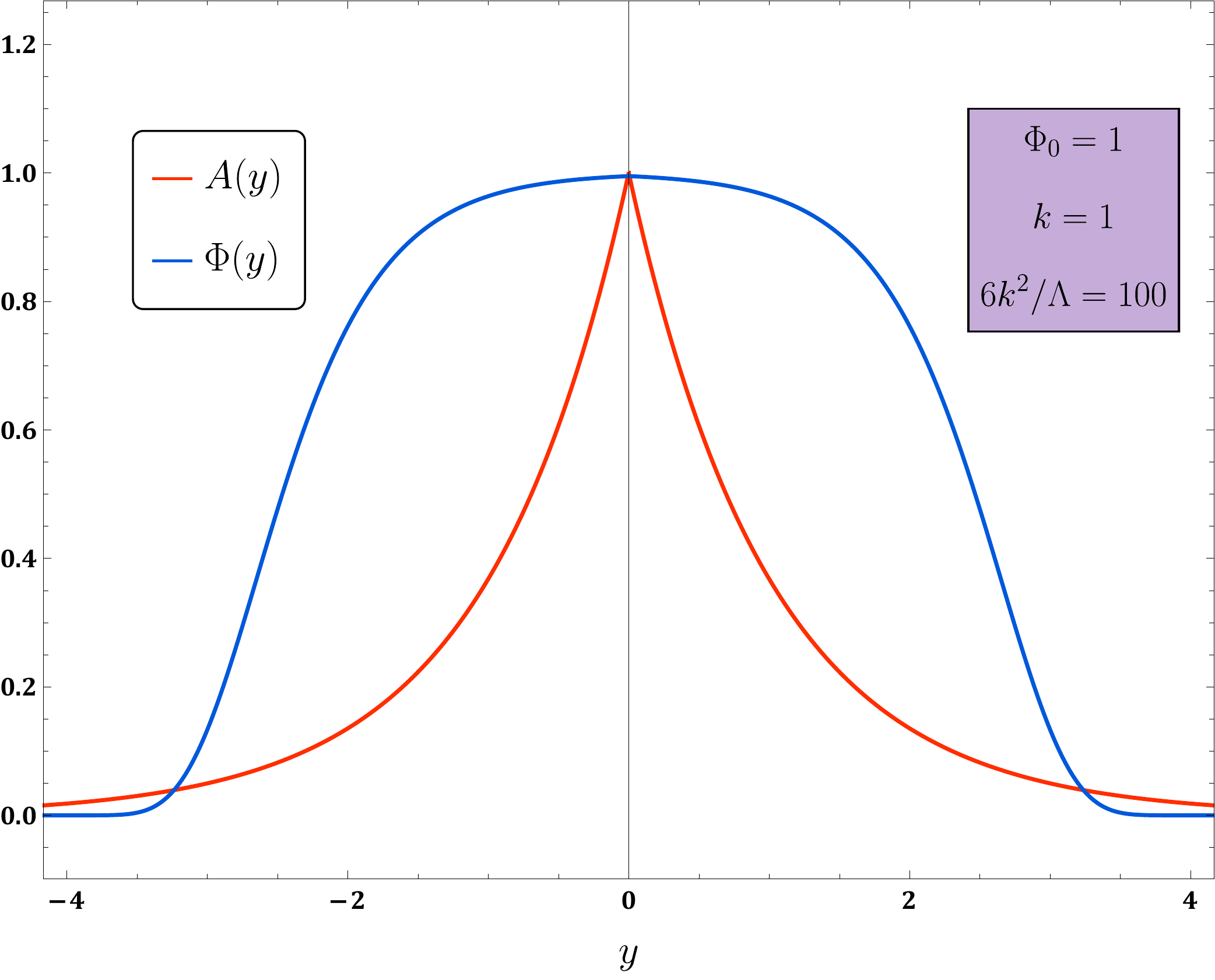}
        \caption{}
        \label{fig3a}
    \end{subfigure}
    \quad
    ~ 
    \begin{subfigure}[b]{0.48\textwidth}
        \includegraphics[width=\textwidth]{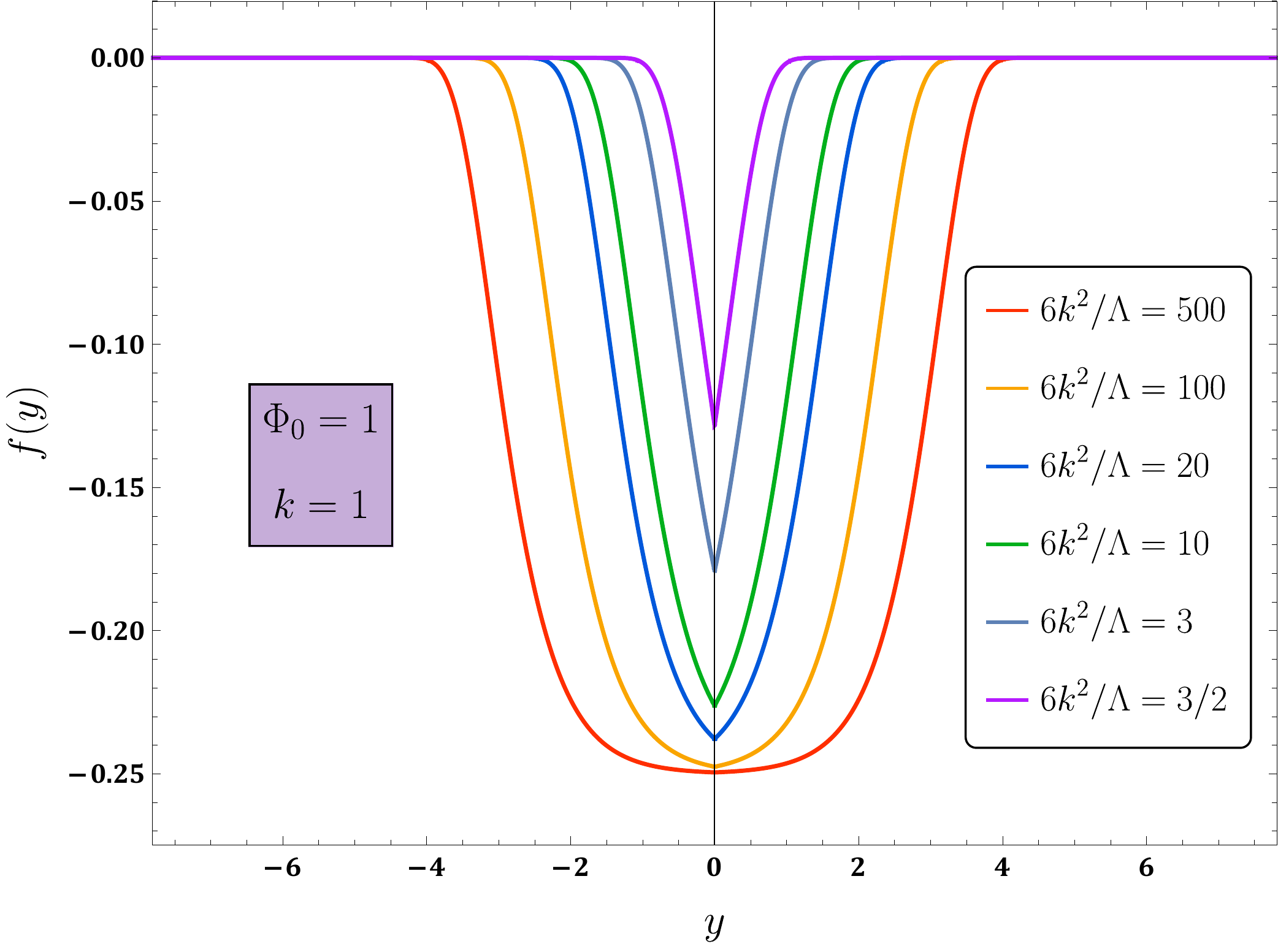}
        \caption{}
        \label{fig3b}
    \end{subfigure}
    ~ 
   \caption{(a) The warp factor $e^{-k |y|}$ and scalar field $\Phi(y)$, and (b)
     the coupling function $f(y)=a \Phi^2(y)$, in terms of the coordinate $y$,
   for $k=1$, $\Phi_0=1$, and $6k^2/\Lambda=3/2,3,10,20,100,500$ (from top to bottom).}
   \label{Aphi_quad}
\end{figure}

We also need to derive the form of the potential $V_B(\Phi)$ of the scalar field in
the bulk. This follows easily from Eq. (\ref{grav-2}) leading to the expression
\beq
V_B(\Phi)=-\Lambda_5 +k^2 \Phi^2\left[\frac{3}{2} + 2 \ln \left(\frac{\Phi}{\Phi_0}\right)
+2  \ln^2 \left(\frac{\Phi}{\Phi_0}\right)\right],
\label{V_quad}
\eeq
or, in terms of the $y$-coordinate,
\beq
V_B(y)=-\Lambda_5 +k^2 \left(\frac{3}{2} -\frac{\Lambda}{6k^2}\,e^{2ky} +
\frac{\Lambda^2}{72k^4}\,e^{4ky} \right) \Phi_0^2 \exp\left(-\frac{\Lambda}{6k^2}\,e^{2ky}\right).
\label{V_quad_y}
\eeq
The bulk potential in principle consists of the negative cosmological-constant term and a
term that is related to the scalar field. For $\Lambda>0$, this second term decreases very
fast exhibiting also a localization around our brane---its profile is shown in Fig. 4(a).
Setting $z=\Lambda e^{2ky}/6k^2$,
one may easily see that the second-order polynomial inside the brackets has no real roots,
and is thus always positive definite. As a result, the second term tends to reduce the
negative bulk cosmological constant, if present, with this effect being more important
close to the brane and negligible far away. In fact, the emergence of a decreasing warp
factor has not been related so far to the presence of $\Lambda_5$.

\begin{figure}[t!]
    \centering
    \begin{subfigure}[b]{0.47\textwidth}
        \includegraphics[width=\textwidth]{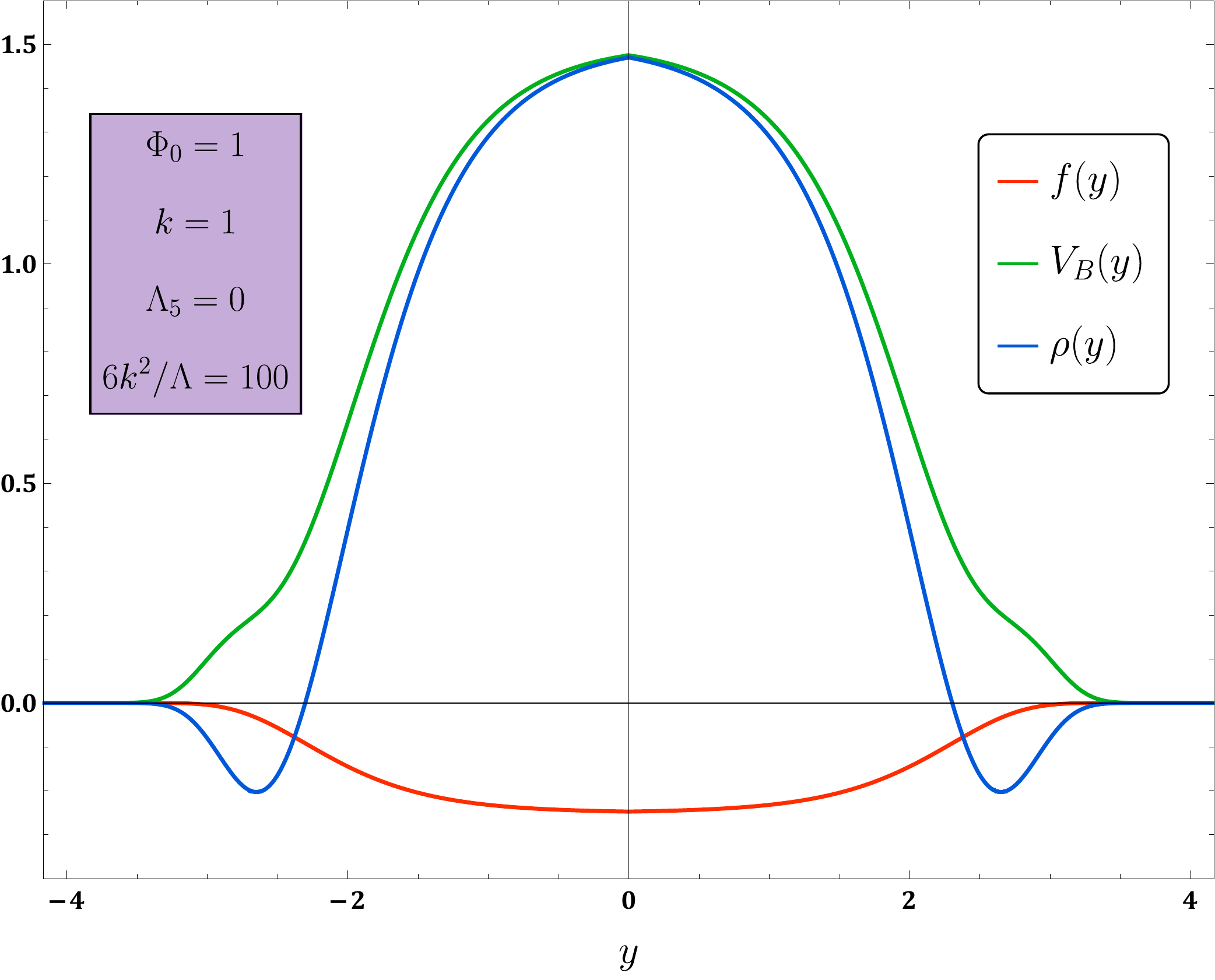}
        \caption{}
        \label{fig4a}
    \end{subfigure}
    \quad
    ~ 
    \begin{subfigure}[b]{0.47\textwidth}
        \includegraphics[width=\textwidth]{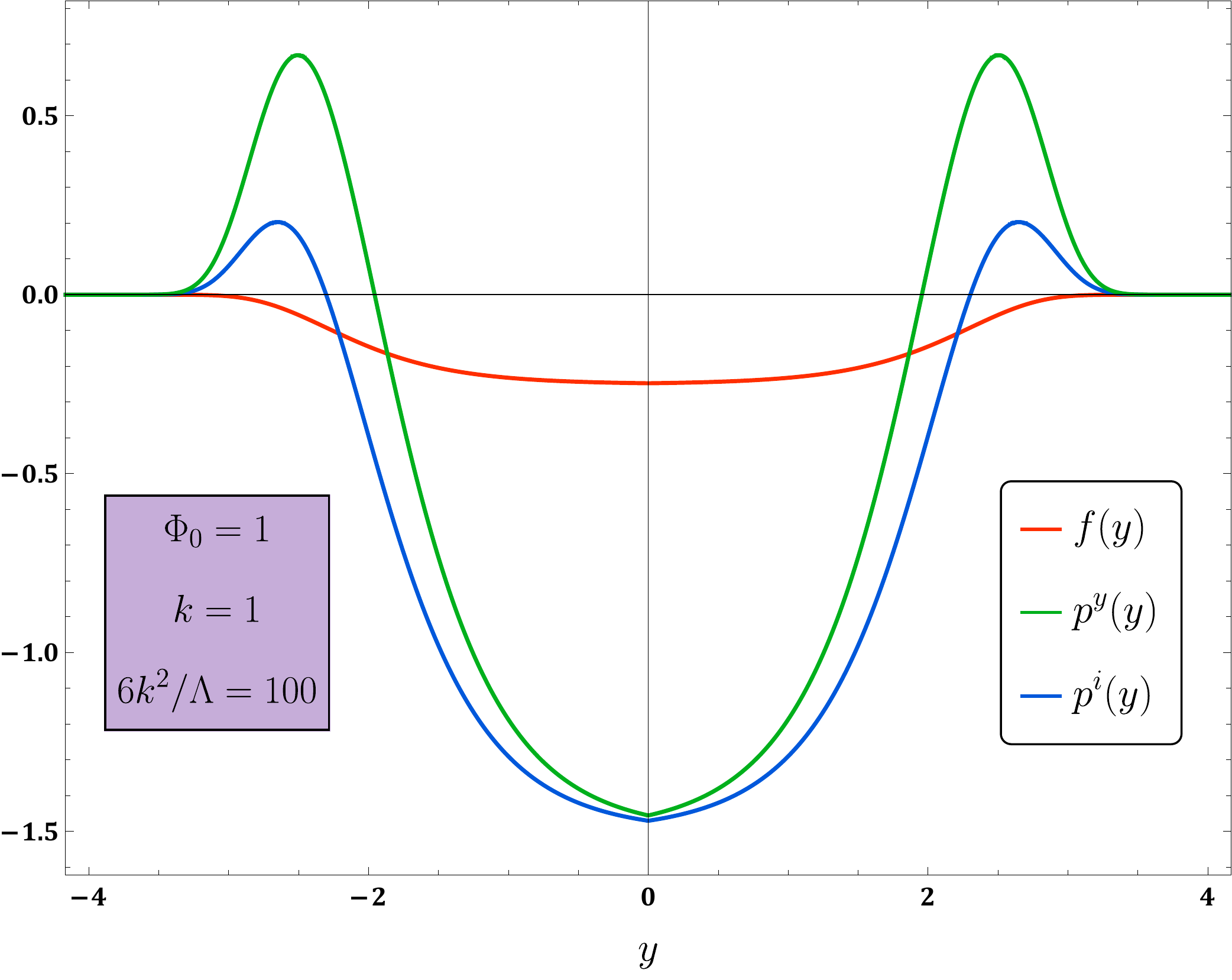}
        \caption{}
        \label{fig4b}
    \end{subfigure}
    ~ 
 \caption{(a) The scalar potential $V_B$ and energy density $\rho$ of the system, and
(b) the pressure components $p^y$ and $p^i$ in terms of the coordinate $y$
(from top to bottom in both plots), for $6k^2/\Lambda=100$, $\Phi_0=1$, $k=1$,
and $\Lambda_5=0$. We also display the coupling function $f$,
for comparison.}
   \label{fig4}
\end{figure}

The components of the energy-momentum tensor may also easily be derived from
Eqs. (\ref{grav_new}) and (\ref{TMN-mixed}) using the solution for the scalar field
and scalar potential. They have the form
\beq
\rho=-p^i=\Lambda_5 +2a \Phi\,(3 A' \Phi' + \Phi'') + V_B(\Phi)
= \frac{3}{2}\,k^2 \Phi^2\left(1-\frac{\Lambda}{6k^2}\,e^{2ky}\right)\,,
\label{T00_quad}
\eeq
\beq
p^y=-\Lambda_5 +\frac{1}{2}\,\Phi'^2 -8a \Phi  A' \Phi' -V_B(\Phi)
= -\frac{3}{2}\,k^2 \Phi^2\left(1-\frac{\Lambda}{3k^2}\,e^{2ky}\right)\,.
\label{Tyy_quad}
\eeq
The form of the energy-momentum components are depicted in Figs. 4(a) and
4(b). The energy density matches the value of the potential at the location
of the brane and decreases slightly faster than the latter away from the brane;
it remains predominantly
positive apart from a small regime at large distances from the brane. The
pressure components exhibit the exact opposite behavior regarding their sign.
Overall, the energy-momentum components resemble those of a {\it positive}
cosmological constant close to our brane, then decrease fast, and finally vanish
at large distances exhibiting a nice localization pattern. We should stress that,
according to our analysis, no negative, bulk cosmological constant needs to be
introduced by hand. It is, in fact, the negative value of the coupling function $f(\Phi)$
that turns the coupling term between the scalar field and the Ricci scalar to a
form of a negative distribution of energy; it is this term then that manages to
support the exponentially falling warp factor even in the absence of a typical
AdS spacetime. 

The presence of the brane, with its nontrivial energy content, introduces once
again discontinuities in the derivatives of the warp factor and scalar field. The
associated junction conditions at $y=0$ have the same form as in 
Eqs. (\ref{jun_linear}). Their explicit forms, however,  are bound to be different
and are given by 
\bea
\frac{k \Phi_0^2}{2}\,e^{-\Lambda/6k^2}\left(3+\frac{\Lambda}{3k^2}\right) &=&
-\sigma -V_b\bigl|_{y=0}\,, \label{jun_quad_ex_1}\\[2mm]
2k\Phi_0\,e^{-\Lambda/12k^2}\left(2+\frac{\Lambda}{6k^2}\right)&=&
-\partial_\Phi V_b\bigl|_{y=0}\,.
\label{jun_lquad_ex_2}
\eea
Since the left-hand sides of the above equations are positive definite, the interaction
term $V_b$ of the scalar field with the brane must be not only nonvanishing but
necessarily negative (with a negative first derivative, too) in order to avoid a negative
brane self-energy $\sigma$. As before, the above conditions may fix the parameters
$k$ and $\Lambda$ while the scalar-field parameters $\Phi_0$ and $V_b$, as well
as $\sigma$, may remain arbitrary.
   
We should, however, stress that this particular solution, being either a black string
or regular in the bulk, cannot constitute a realistic model due to the negative sign
of the coupling function $f(\Phi)$. This sign will be carried over to the four-dimensional
effective theory leading to antigravity on the brane. Indeed, working as in the previous
subsection and isolating the coefficient of $R^{(4)}$ in the action, we arrive at the
relation
\beq
\frac{1}{\kappa_4^2} = -\frac{\Phi_0^2}{2}\,\int_{0}^{\infty} dy\,e^{-2 k y}\,
\exp\left(-\frac{\Lambda}{6k^2}\,e^{2ky}\right) = -\frac{\Phi_0^2}{4k}
\left(e^{-\Lambda/6k^2}-\frac{\Lambda}{3k}\,I\right),
\label{effG_1_quad}
\eeq
where
\beq
I \equiv \int_{0}^{\infty} dy\,\exp\left(-\frac{\Lambda}{12k^2}\,e^{2ky}\right).
\label{I-def}
\eeq
The integral $I$ may be computed numerically and yields a finite result; therefore, there
is no need for the introduction of a second brane in this model.\footnote{A similar
analysis to that of Sec. 3.1, but simpler due to the absence of the second brane, leads
to the derivation of the effective cosmological constant on the brane, which once again
comes out to be $\Lambda_4=\Lambda/\kappa^2_4$.} Nevertheless, the value
of the effective gravitational constant $\kappa^2_4$ turns out to be negative---this 
becomes clear if one looks at the middle part of Eq. (\ref{effG_1_quad}), where a 
negative coefficient multiplies a positive-definite integral. This result is catastrophic, and
therefore, the model is not a viable one. Its emergence, however, reveals two facts:
(i) that antigravitating solutions in the context of the theory (\ref{action}) are
somehow associated with the positive cosmological constant on the brane since two
such solutions have emerged for two different choices of the coupling function, and
(ii) that, when $M \neq 0$, the theory of a nonminimally coupled scalar field to
gravity gives rise to yet another undesired black-string solution rather than a physically
motivated, and long sought-for, localized black-hole solution.


\section{A Theoretical Argument}
\label{Theoretical}

In the previous section, we have constructed explicit solutions that emerge 
from the five-dimensional field equations, and we describe a four-dimensional
Schwarzschild--de Sitter background on the brane. From the bulk point of view,
these solutions describe either black strings, if $M \neq 0$, or regular,
maximally symmetric solutions over the whole bulk apart from its
boundary at $y \rightarrow \infty$, if $M=0$---a second brane could easily shield
the boundary singularity creating two-brane models with a compact fifth
dimension. In both cases, however, the bulk solution is characterized,
either globally or over particular regimes, by a negative coupling function
$f(\Phi)$ that leads to an antigravitating theory. In this section, we
examine from the mathematical point of view why the emergence of such
solutions is possible in the context of the given theory, and why they
do so particularly for the physically motivated case of a positive cosmological
constant on the brane.

The analysis will focus on the gravitational equation \eqref{grav-1}. By
employing the relations
\beq\label{321}
\pa_y f= \Phi'\,\pa_\Phi f\,, \qquad 
\pa_y^2 f=\Phi'^2 \,\pa_\Phi^2 f+\Phi''\,\pa_\Phi  f\,,
\eeq
as well as the expressions $A(y)=-k|y|$ and $m(r)=M+\Lambda r^3/6$, Eq. \eqref{grav-1}
is written as
\eq$\label{grav1-v0}
\Lambda\ e^{2k|y|} f=-\Phi'^2-\pa_y^2 f-k\, sgn(y)\, \pa_y f\,.$
We assume once again the existence of the $Z_2$-symmetry in the bulk and restrict
our analysis to the positive $y$-regime for simplicity. Consequently, we write
\eq$\label{grav1-v1}
\Phi'^2=-\pa_y^2 f-k\pa_y f-\Lambda e^{2ky} f\,.$
The first derivative of the scalar field $\Phi'$ may vanish at particular
values of the coordinate $y$ but is assumed to be in general nonvanishing
to allow for a nontrivial scalar field in the bulk. Also, both functions $f=f(y)$ and
$\Phi=\Phi(y)$ should be real in their whole domain. Therefore, both sides of
Eq. (\ref{grav1-v1}) should be non-negative, which finally leads to the constraint
\beq
\label{grav1-v2}
\pa_y^2 f+ k\pa_y f + \Lambda e^{2ky} f \leq 0\,.
\eeq
The above constraint should be satisfied for every solution of the field
equations (\ref{eq-mass})--(\ref{grav-2}), including the ones presented in
Secs. 3.1 and 3.2. These were characterized by $\Lambda>0$, in which case
the combination $\Lambda e^{2ky}$, appearing in the last term of the above
expression, diverges to $+\infty$ at the boundary of spacetime. But there,
the coupling function $f(y)$ is negative for both solutions, and this renders
the last, dominant term smaller than zero as the constraint demands. Also,
for all other values of $y$, one may easily check that the profiles of the
function $f(y)$ found in Secs. 3 and 4 always satisfy the constraint
(\ref{grav1-v2}).

In what follows, we investigate whether physically acceptable
solutions with $f(y)>0$ may arise in the case where $\Lambda$ is also positive. 
To this, we will add the demand that the components of the energy-momentum
tensor may be localized close to the brane, and are certainly nondiverging
at the boundary of spacetime. These may be written as
\bea
\rho=-p^i &=& \frac{1}{2}\,\Phi'^2 +V_B(\Phi) + \Lambda_5+ 3A' \partial_y f
+ \partial_y^2 f\,, \label{rho-general} \\[1mm]
p^y &=&  \frac{1}{2}\,\Phi'^2 -V_B(\Phi) -\Lambda_5- 4A' \partial_y f\,.
\label{py-general}
\eea
From Eq. (\ref{grav-2}), one may solve for the general form of the scalar
potential to find
\beq
V_B(\Phi)=-\Lambda_5 -\frac{1}{2}\,\Phi'^2 - 3A' \partial_y f - \partial_y^2 f
-f\,(6k^2-\Lambda\,e^{2ky})\,. \label{potential-general}
\eeq
Employing the above into the expressions (\ref{rho-general}) and (\ref{py-general}),
together with Eq. (\ref{grav1-v1}), the energy-momentum tensor components
simplify to
\bea
\rho=-p^i &=& -6k^2 f(\Phi)+ f(\Phi)\,\Lambda e^{2ky}\,, \label{rho-final} \\[2mm]
p^y &=& 6k^2 f(\Phi)-2 f(\Phi)\,\Lambda e^{2ky}\,.
\label{py-final}
\eea
We observe that all components contain the diverging combination $\Lambda e^{2ky}$.
Therefore, we should demand the vanishing of the coupling function $f(\Phi)$ at the
boundary of spacetime at least as fast as $e^{-2ky}$. 

We will consider the most general such form, namely $f(y)=A e^{-\sum_{n=1}^N b_n y^n}$,
where $A$ and $b_n$ are arbitrary constants, and $N$ is a positive integer. The
first and second derivatives of $f(y)$ are found to be
\bea
\pa_y f &=&-A e^{-\sum_{n=1}^N b_n y^n}\left(\sum_{n=1}^N b_n n y^{n-1}
\right), \nonumber \\[2mm]
\pa_y^2 f&=&A e^{-\sum_{n=1}^N b_n y^n} \left[\left(\sum_{n=1}^N b_n n y^{n-1} \right)^2 -\sum_{n=1}^N b_n n(n-1) y^{n-2}\right]. \label{df_exp}
\eea
Both quantities quickly tend to zero which ensures the finiteness of the scalar potential
(\ref{potential-general}). Then, the inequality constraint of Eq. (\ref{grav1-v2}) reads
\gat$
f(\Phi)\left[\left(\sum_{n=1}^N b_n n y^{n-1} \right)^2 -\sum_{n=1}^N b_n n(n-1) y^{n-2}
-k\sum_{n=1}^N b_n n y^{n-1} +\Lambda e^{2ky}\right] \leq 0\,. \label{con-final}$
Since $f(\Phi)$ is demanded to be everywhere positive, it is the expression inside the
square brackets that needs to be negative definite. For $N=1$, the latter reduces to
$b_1 (b_1-k) +\Lambda e^{2ky}$; but this, for $\Lambda>0$, is always positive definite
since $b_1\geq 2k$ according to the argument below Eqs. (\ref{rho-final}) and (\ref{py-final}).
For $N>1$, as $y$ increases away from the brane, the first and last terms are clearly
the dominant ones in Eq. (\ref{con-final}); but these are again positive definite. Therefore,
in all cases the constraint (\ref{grav1-v2}) is violated either over the entire $y$-regime
(as in the case studied in Sec. 4) or at a distance from the brane (as in the case
studied in Sec. 3).

\par Conclusively, we have demonstrated that, for a function $f(\Phi)$ positive and decreasing
at large distances from our brane---assumptions that guarantee the correct sign of the
gravitational force and the localization of the energy-momentum tensor in the bulk---
no viable solutions arise in the context of the theory (\ref{action}) when $\Lambda>0$
on our brane. On the other hand, for $\Lambda$ either zero or negative, solutions with
$f>0$ are much easier to arise.\footnote{In fact, a static braneworld solution with $M=0$
and $\Lambda=0$ on our brane was presented in \cite{Bogdanos1} where a quadratic coupling
function $f(\Phi)=1-\xi \Phi^2$ between the scalar field and the Ricci scalar was considered.}
For example, for $\Lambda=-|\Lambda|<0$,  Eq. (\ref{grav1-v2}) is now written as
\eq$\label{con-AdS}
\pa_y^2 f+k \pa_y f - f |\Lambda|\,e^{2ky}\leq 0\,.$
One may readily see that this constraint is much easier to satisfy: for $f(\Phi)$ positive
and decreasing, the second and third terms are already negative definite. For instance,
the choice considered above for $f(\Phi)$, namely $f(y)=A e^{-\sum_{n=1}^N b_n y^n}$,
satisfies the constraint (\ref{con-AdS}) over the entire $y$-regime for appropriate choices
of the parameters. A detailed analysis on the emergence of legitimate solutions in the
context of the theory (\ref{action}) with a Minkowski or anti de Sitter background on our
brane will be given elsewhere \cite{KNP2}.



\section{Conclusions}

Motivated by the results of previous works \cite{KPZ, KPP2}, where despite intensive
efforts regular, localized-on-the-brane black hole solutions were not found
in the context of a theory with a scalar field nonminimally coupled to gravity,
in this work we have focused on the derivation and study of the properties of
black-string solutions that, in contrast, seem to emerge quite naturally in the
context of the same theory. To this end, we have retained the Vaidya form
of the spacetime line element, which on the brane leads to a Schwarzschild black
hole while in the bulk produces solutions with the minimum number of
spacetime singularities. We have in addition allowed for an arbitrary mass
function $m(r)$ in an effort to accommodate, if possible, solutions with a more
general profile including an (anti)--de Sitter or Reissner-Nordstrom type of 
background.

The integration of an appropriate rearrangement of the equations of motion 
has allowed us to uniquely determine the form of the mass function, namely
$m(r)=M + \Lambda r^3/6$.  Performing an inverse coordinate transformation
on the brane, we readily identified the parameter $M$ with the black-hole
mass and the parameter $\Lambda$ as the product $\kappa_4^2 \Lambda_4$,
where $\Lambda_4$ is the four-dimensional cosmological constant on the brane.
As a result, the brane background assumes the form of a Schwarzschild--(anti)--de
Sitter spacetime. As the expressions of the five-dimensional curvature invariants
reveal, these solutions may have a dual description from the bulk point of view:
they may describe either black strings, if $M \neq 0$, or braneworld
maximally symmetric solutions, if $M=0$. 

The properties of these five-dimensional solutions strongly depend on the 
form of the nonminimal coupling function $f(\Phi)$ between the scalar field and
the five-dimensional scalar curvature. We have considered two simple choices for 
$f(\Phi)$, a linear one and a quadratic one in terms of the scalar field. For a linear
coupling function, the 
scalar field is found to increase exponentially away from the brane and to drive
the coupling function to negative values at a distance from the brane. When
$6k^2/\Lambda>1$, there is always a positive-value regime for $f(\Phi)$ close
to our brane while the antigravitating regime, with $f(\Phi)<0$, is pushed away
from our brane as the value of $6k^2/\Lambda$ gradually increases. For fairly
large values of $6k^2/\Lambda$, i.e. for a large warping factor $k$ or a small
cosmological constant on our brane, the profile of the coupling function exhibits
a wide plateau around our brane. When $6k^2/\Lambda \simeq 125$, this plateau
is centered
around the value of unity, and, therefore, the theory mimics a five-dimensional
scalar-tensor theory with a minimally coupled scalar field and normal gravity
around our brane---the antigravitating regime is, however, still lurking at the
boundaries of the extra dimension. The latter may be cut short or altogether 
removed from the theory by adding a second brane; this is also necessary in
order to obtain a finite four-dimensional gravitational scale, as we have explicitly
demonstrated. The antigravitating regime is also characterized by a diverging
scalar field that results in the divergence of the energy-momentum tensor components,
too. However, after the introduction of the second brane at a finite distance from
the first, all energy-momentum tensor components are well behaved. In fact, the
energy density takes on an almost constant, negative value around our brane,
thus mimicking a bulk cosmological constant (which, in this case, is redundant)
and supporting a Randall-Sundrum warp factor.

For a quadratic coupling function $f(\Phi)$, the scalar field is found to be everywhere
finite and, in fact, to exhibit a localization around our brane---the same behavior
is exhibited by all the energy-momentum tensor components. The four-dimensional
gravitational scale comes out to be finite; therefore, in this case there is no reason
to introduce a second brane. The warp factor takes a form identical to the one in
the Randall-Sundrum model even in the absence of a negative, bulk cosmological
constant and for positive values of the energy-momentum tensor around our brane.
What, in fact, creates the anti--de Sitter spacetime in the bulk and supports the
exponentially decreasing warp factor is the coupling itself between the scalar field
and the bulk scalar curvature, which is everywhere negative. This, of course, leads
to an antigravitating theory over the whole spacetime and eventually to an
unphysical gravitational theory on our brane. This model, being far from a 
realistic theory, is nevertheless a characteristic example of the variety of solutions
that may arise in braneworlds; more specifically, it underlines the easiness with
which unphysical black-string solutions (in the case where $M \neq 0$) emerge
in contrast to the physically motivated localized black-hole solutions. 

The discussion of the second model with the quadratic coupling function also served
another purpose: together with the first one with a linear $f(\Phi)$, they were
both derived under the assumption of a positive cosmological constant on our brane. 
Also, both models were characterized, either globally or over particular regimes, by a
negative coupling function $f(\Phi)$ that led to an antigravitating theory. In order
to investigate the potential connection between a Schwarzschild--de Sitter spacetime on our
brane and an antigravitating regime in the bulk, in Sec. 5, we examined from
the mathematical point of view why the field equations in the present theory seem to favor
the emergence of these solutions. By turning a particular combination of the field
equations into a constraint relating solely the coupling function, its derivatives,
and the effective cosmological constant, we demonstrated that, for $\Lambda>0$,
this constraint is impossible to satisfy for $f(\Phi)$ also positive for the entire extra
dimension. 
Therefore, in this class of theories, with a nonminimally coupled scalar field and
a general coupling function, the emergence of an effective four-dimensional theory
on our brane with a positive cosmological constant is always accompanied by
a problematic antigravitating regime in the five-dimensional bulk. 

The aforementioned conclusion opens the way for the derivation of solutions
with normal gravity in the case of either a Minkowski or anti--de Sitter spacetime
on our brane.  Although less physically motivated, it would still be of interest 
to investigate whether a scalar-tensor theory in the bulk could support a
solution (either a black string or a regular one) with a decaying warp factor
but without the need for a constant distribution of a negative energy density
in the higher-dimensional spacetime. Another question we also hope to come
back to is that of the stability of the solutions, the ones derived here and the
ones soon to be presented, under perturbations. Note that the nonminimal
coupling of the scalar field to the gravitational field makes the stability analysis
highly nontrivial compared to the pure gravitational solutions derived in the
literature (see, for example, \cite{Farakos1}). Given the fact that usually
black-string solutions suffer from the Gregory-Laflamme instability \cite{GL, RuthGL}---which
up to now has served as a method to get rid of unphysical solutions,  
it will be extremely interesting to see whether the presence of the scalar field and the
sign of its coupling function $f(\Phi)$ affects in any way the stability of these solutions.


{\bf Acknowledgement.}
The research of N.P. was implemented under the ``Strengthening Post-Doctoral
Research" scholarship program (grant no 2016-050-0503-7626) by the Hellenic
State Scholarship Foundation as part of the Operational Program ``Human
Resources Development Program, Education and Lifelong Learning", co-financed by
the European Social Fund-ESF and the Greek government.

\appendix

\numberwithin{equation}{section}	


\section{Curvature Invariant Quantities}
\label{App-Invar}

Employing the expression of the line element (\ref{metric}), one may compute the
scalar curvature invariant quantities. These have the form
\eq$\label{4}
R=-8 A''-20 A'^2+\frac{2 e^{-2 A} \left(r \pa_r^2m+2 \pa_rm\right)}{r^2}\,,$
\begin{align}\label{5}
R_{MN}R^{MN}=2e^{-4 A} \left[e^{2 A} \left(A''+4 A'^2\right)-\frac{\pa_r^2m}{r}\right]^2&+2\frac{e^{-4 A} \left[r^2 e^{2 A}\left(A''+4 A'^2\right)-2 \pa_rm\right]^2}{r^4}\nonum\\
&+16 \left(A''+A'^2\right)^2,
\end{align}
\begin{align}\label{6}
R_{MNKL}R^{MNKL}=&-\frac{8 e^{-2 A} A'^2 \left(r \pa_r^2m+2 \pa_rm\right)}{r^2}+40 A'^4+16 A'' \left(A''+2 A'^2\right)\nonum\\[2mm]
&+4 e^{-4 A}\left[\frac{(\pa_r^2m)^2}{r^2}+\frac{4\left[2 (\pa_rm)^2+ \left(m-r \pa_rm\right) \pa_r^2m\right]}{r^4}+\frac{4 (3 m^2-4 r m \pa_rm)}{r^6}\right],
\end{align}
and may be used for the geometric characterization of the solutions derived from the field
equations.


\section{Independent Field Equations}
\label{App-Indep}

Here, we will demonstrate that the three field equations (\ref{grav-1})--(\ref{phi-eq})
are not all  independent. To this end, we substitute the mass function $m(r)=M+\Lambda r^3/6$
into Eq. (\ref{grav-1}); as shown in Sec. \ref{Theoretical}, the latter may then be
brought to the form 
\gat$
\label{311}
\Phi'^2=-f(3A''+\Lambda e^{-2A})+A'\pa_y f-\pa_y^2f\,.$
Taking the derivative of both sides with respect to $y$, we obtain
\gat$ \label{312}
2 \Phi'\,\Phi''=-f(3A'''-2\Lambda A'e^{-2A})-\pa_y f(2A''+\Lambda e^{-2A})
+A'\pa_y^2 f-\pa_y^3 f\,.$

Next, we consider Eq. (\ref{grav-2}) which we solve for the potential $V$ to find
\gat$ \label{313}
V=-\Lambda_5-\frac{1}{2}\,\Phi'^2-f(6A'^2+3A''-\Lambda e^{-2A})-3A'\pa_y f-\pa_y^2 f\,.$
If we take again the derivative with respect to $y$, we arrive at the result
\gat$
\hspace*{-3cm}\pa_yV=-\Phi'\,\Phi''-f(12A'A''+3A'''+2\Lambda A'e^{-2A})\nonum\\[1mm] \label{314}
\hspace{4.8cm} -\pa_yf(6A'^2+6A''-\Lambda e^{-2A}) -3A'\pa_y^2f-\pa_y^3f\,.$
We now use the above expression in the scalar-field equation (\ref{phi-eq}) after multiplying
first the latter by $\Phi'$; we eventually obtain
\gat$ \label{315}
2\Phi'\,\Phi''=-f(3A''-2\Lambda A'e^{-2A})-\pa_yf(2A''+\Lambda e^{-2A})+A'\pa_y^2f-\pa_y^3f\,.$

\par We see that Eqs. \eqref{312} and \eqref{315} are identical, which means that the 
three field equations from which these equations were derived are not independent. We are thus
entitled to keep only two of them in our analysis and to ignore the third one. 


\section{Inverse Generalized Vaidya Transformation}
\label{App-Vaidya}

Starting from the projected-on-the-brane line element (\ref{metric-4D}), which we will write
for simplicity as 
\eq$\label{vtos1}
ds^2=-\left(1-\frac{2m(r)}{r}\right)dv^2+2dvdr+r^2(d\theta^2+\sin^2\theta\,d\varphi^2)\,,$
where $m(r)=M+\Lambda r^3/6$, we will seek to determine the coordinate transformation of the
Vaidya time-variable $v$, if existent, that will bring the aforementioned line element to a
diagonal, Schwarzschild-like form
\beq
ds^2=-f(r)\,dt^2+\frac{dr^2}{f(r)}+r^2(d\theta^2+\sin^2\theta\,d\varphi^2)\,.
\label{diagonal}
\eeq

We will consider the following general transformation:
\eq$\label{vtos3}
v=h(t,r) \,\Rightarrow\, dv=\pa_th\ dt+\pa_rh\ dr\,.$
Substituting the above expression of $dv$ into Eq. \eqref{vtos1}, we obtain
\bal$\label{vtos5}
ds^2&=-\left(1-\frac{2m(r)}{r}\right)(\pa_th)^2dt^2+\left[-\left(1-\frac{2m(r)}{r}\right)(\pa_rh)^2+2\pa_rh\right]dr^2\nonum\\
&\hspace{12.27em}+2\pa_th\left[-\left(1-\frac{2m(r)}{r}\right)\pa_rh+1\right] dtdr+r^2d\Omega^2\,.$
We now demand the vanishing of the off-diagonal term in Eq. (\ref{vtos5}): for 
$\partial_t h \neq 0$, this leads to the constraint
\beq
\partial_r h=  \left(1-\frac{2m(r)}{r}\right)^{-1}\,.
\eeq
Provided that the above holds, the coefficient of $dr^2$ in Eq. (\ref{vtos5}) reduces to
$\partial_r h$, and therefore
\eq$\label{vtos6}
f(r)=\frac{1}{\partial_r h}=\left(1-\frac{2m(r)}{r}\right).$
Comparing finally the coefficients of $dt^2$ in Eqs. \eqref{diagonal} and \eqref{vtos5}, we
conclude that $\pa_th$ must be equal to unity. Therefore, if the coordinate transformation
$v= t+g(r)$ is applied to the line element (\ref{vtos1}), the latter takes indeed the diagonal form\\
$ds^2=-\left(1-\frac{2M}{r} -\frac{\Lambda r^2}{3}\right) dt^2+
\left(1-\frac{2M}{r} -\frac{\Lambda r^2}{3}\right)^{-1} dr^2+r^2(d\theta^2+
\sin^2\theta\,d\varphi^2)$ that describes a four-dimensional Schwarzschild--(anti)--de Sitter background depending
on the sign of the parameter $\Lambda$, which turns out to be proportional to the
cosmological constant on the brane. 

\par To complete the analysis, we need to determine the value of the function $g(r)$
through the integral
\beq
g(r)= \int \frac{dr}{f(r)}=\int \frac{dr}{1-\frac{2M}{r}-\frac{\Lambda r^2}{3}}\,.
\label{integral_g}
\eeq
Evaluating the above integral amounts to calculating the tortoise coordinate for the
specific black-hole background.
The steps of the evaluation depend on the sign of the parameter $\Lambda$. Let us start with
the case $\Lambda>0$, where the four-dimensional background is a Schwarzschild--de Sitter
one. The function $f(r)$ has, in the most general case, two real, positive roots $r_h$ and
$r_c$ corresponding to the black-hole and cosmological horizon, respectively. Then, the aforementioned integral becomes
\beq
g(r)= \frac{3}{\Lambda}\,\int \frac{r\,dr}{-r^3 +3r/\Lambda -6M/\Lambda}=
\frac{3}{\Lambda}\,\int \frac{r\,dr}{(r-r_h)\,(r_c-r)\,(r+r_c+r_h)}\,,
\label{integral_g_dS}
\eeq
where the two horizons satisfy the relations
\beq
(r_c+r_h)^2-r_cr_h=\frac{3}{\Lambda}\,, \qquad (r_c+r_h)\,r_c r_h =\frac{6M}{\Lambda}\,.
\eeq
Splitting the fraction in the integral (\ref{integral_g_dS}) into three separate ones and
performing the corresponding integrations, we arrive at the result \cite{KGB}
\beq
g(r)=\frac{ r_h\,\ln (r-r_h)}{1-\Lambda r_h^2} + \frac{ r_c\,\ln (r_c-r)}{1-\Lambda r_c^2}
-\frac{ (r_c+r_h)\,\ln (r+r_c+r_h)}{1-\Lambda (r_c+r_h)^2} + C_1\,,
\label{g_dS}
\eeq
where $C_1$ is an arbitrary integration constant.

If, on the other hand, $\Lambda=-|\Lambda|<0$, then the background on the brane is of a 
Schwarzschild--anti--de Sitter type. The function $f(r)$ vanishes only at $r=r_h$, i.e. at the
location of the black-hole horizon. We then write:
\beq
g(r)= \int \frac{r\,dr}{\frac{|\Lambda| r^3}{3} +r -2M}=
\frac{3}{|\Lambda|}\,\int \frac{r\,dr}{(r-r_h)\,(r^2+r_h\,r+\beta)}\,,
\label{integral_g_Anti}
\eeq
where $\beta=6M/|\Lambda| r_h$. Note that the quadratic polynomial $r^2+r_h\,r+\beta$ 
has no real, positive roots.  We then split the fraction inside the integral into two
separate ones of the form
\beq
\frac{1}{(r-r_h)\,(r^2+r_h\,r+\beta)} =\frac{A}{r-r_h} +
\frac{B r + D}{r^2+r_h\,r+\beta}\,,
\label{split_Anti}
\eeq
where
\beq
A=\frac{1}{2 r_h^2+\beta}\,, \qquad B=-A\,, \qquad
D=-2r_h A\,.
\eeq
Substituting Eq. (\ref{split_Anti}) into Eq. (\ref{integral_g_Anti}) and applying standard
integration techniques, we finally arrive at the result
\beq
g(r)=\frac{3}{|\Lambda|(2r_h^2+\beta)}\left[r_h\ln\left(\frac{r-r_h}{\sqrt{r^2+r_h\,r+\beta}}\right)    
+ \frac{r_h^2+2\beta}{\sqrt{4\beta-r_h^2}}
\arctan\left[\frac{2r+r_h}{\sqrt{4\beta-r_h^2}}\right]\right]+C_2\,,
\label{f_final_Anti}
\eeq
where $C_2$ is again an arbitrary integration constant and the horizon radius may be
expressed as 
\beq
\label{horizon_Anti}
r_h=\frac{1}{(-3\Lambda^2 M +\sqrt{9 \Lambda^4 M^2+|\Lambda|^3})^{1/3}}-
\frac{(-3\Lambda^2 M +\sqrt{9 \Lambda^4 M^2+|\Lambda|^3})^{1/3}}{|\Lambda|}\,.
\eeq

\bibliographystyle{unsrt}
\bibliography{Bibliography}
\addcontentsline{toc}{chapter}{\numberline{}References}

\end{document}